\documentclass[useAMS,usenatbib]{mn2e}
\usepackage{natbib}
\usepackage{graphicx}
\usepackage{subfigure}
\usepackage{multirow}
\usepackage{amsmath}
\usepackage{amssymb}
\usepackage{epstopdf}
\usepackage{rotating}
\usepackage{soul}

\title[Massive black hole seeds]{
Cosmological simulations of massive black hole seeds: predictions for next
generation electromagnetic and gravitational wave observations}
\author[DeGraf \& Sijacki]  {C. DeGraf$^{1}$, D. Sijacki$^{1}$ \\
{$^1$} {Institute of Astronomy and Kavli Institute for Cosmology, University of Cambridge, Madingley Road, Cambridge CB3 0HA, UK} 
}
\def\simgt{\lower.5ex\hbox{$\; \buildrel > \over \sim \;$}}

\begin{document}

\date{Submitted to MNRAS}
\pubyear{2019}

\maketitle
\begin{abstract}
We study how statistical properties of supermassive black holes depend on the
frequency and 
conditions for massive seed formation in cosmological simulations of structure
formation. We develop a novel method to re-calculate detailed growth histories
and merger trees of black holes within the framework of the Illustris
simulation for several seed formation models, including a physically motivated
model where black hole seeds only form in progenitor galaxies that conform to
the conditions for direct collapse black hole formation. While all seed models
considered here are in a broad agreement with present observational
constraints on black hole populations from optical, UV and X-ray studies, we
find they lead to widely different black hole number densities and halo
occupation fractions which are currently observationally unconstrained. In
terms of future electromagnetic spectrum observations, the faint-end quasar
luminosity function and the low
mass-end black hole-host galaxy scaling relations are very sensitive to the
specific massive seed
prescription. Specifically, the direct collapse model exhibits a seeding efficiency which decreases rapidly with cosmic time and produces much fewer black holes in low mass galaxies, in contrast to the original Illustris simulation. We further find that the total black hole merger rate varies by more than one order of magnitude for different seed models, with the redshift evolution of the chirp mass changing as well. Supermassive black hole merger detections with LISA and International Pulsar Timing Array may hence provide the most direct means of constraining massive black hole seed formation in the early Universe.

\end{abstract}

\section{Introduction}
\label{sec:intro}

It is widely understood that supermassive black holes (SMBHs) are found at the centre of massive galaxies \citep{KormendyRichstone1995}, and that properties of the host galaxy strongly correlate with black hole mass, suggesting a causal link between them \citep[e.g.][]{Magorrian1998, Gebhardt2000, Graham2001,Ferrarese2002, Tremaine2002, HaringRix2004, Gultekin2009, McConnellMa2013, KormendyHo2013}.  

Observations have also shown that high-redshift quasars are found at $z \sim
6$ \citep[e.g.][]{Fan2006, Jiang2009} and even $z > 7$, with black hole masses
estimated at $\sim 10^9 \mathrm{M_\odot}$ \citep{Mortlock2011, Banados2018}.  Reaching
such high masses at such an early cosmic time provides a challenge to the
theoretical models of supermassive black hole growth, and implies the
possibility of the formation of sufficiently massive seeds followed by growth
capable of reaching $10^9 \mathrm{M_\odot}$ in less than $10^9$~yrs of cosmic time.
Cosmological simulations have shown that extended periods of Eddington growth
can be sustained by high density gas flows in very over-dense regions of the
Universe, providing an almost constant fuel source for the accreting black
hole, such that $\sim 10^9 \mathrm{M_\odot}$ black holes are produced by $z \sim 7$
when starting with seeds of $\sim 10^5 \mathrm{M_\odot}$ \citep[e.g.][]{Sijacki2009,
  DiMatteo2012, Costa2014, Curtis2016}.  

The initial formation of supermassive black holes seeds, however, is an open
question, with many different mechanisms proposed which may lead to the
formation of light (stellar), intermediate and massive seeds \citep[see the
  seminal paper by][]{Rees1984}. Light seed formation has been proposed in the context of remnants of Population III (PopIII) stars \citep[e.g.][]{MadauRees2001,Volonteri2003}.  After the first generation of stars forms in pristine haloes at very high redshift, stars with masses from $\sim 25-140 \mathrm{M_\odot}$ and $\gtrsim 260 \mathrm{M_\odot}$ are expected to collapse to form a black hole \citep{Heger2003, HegerWoosley2010, WhalenFryer2012, Karlsson2013}, producing light seeds with $M_{\rm{BH,seed}}$ on the order of half the stellar mass \citep[though uncertainty in the PopIII initial mass function could produce $M_{\rm{BH,seed}} \sim 10-1000 \: \mathrm{M_\odot}$, e.g.][]{Fryer2001,Hirano2014}.  While seed formation from PopIII stellar remnants is potentially common (though with some uncertainty in the mass scales), the low-mass nature of these seeds would require substantially more growth and possibly super-Eddington accretion to reach $\sim 10^9 \: \mathrm{M_\odot}$ by $z \sim 7$ making them potentially problematic as the progenitors of bright z $\gtrsim 6$ quasars.

At an intermediate seed mass scale, it is proposed that black holes may form
in dense nuclear star clusters (NSC), where runaway stellar collisions can
produce a very massive star which can then collapse to an intermediate mass
black hole, on the order of $M_{\rm{BH,seed}} \sim 10^3 \: \mathrm{M_\odot}$
\citep[e.g.][]{BegelmanRees1978,PortegiesZwart2002,Freitag2006a,
  Freitag2006b,Omukai2008,DevecchiVolonteri2009,Katz2015}.  The higher initial
mass compared to PopIII remnants makes the prospects of reaching
bright quasar masses at $z \sim 6$ more viable but not without a challenge of
sustained and very high growth rate.

Massive seed formation is generally proposed to occur in haloes with
$T_{\rm{vir}} > 10^4$ K containing (almost) pristine gas which can potentially
collapse directly to a black hole without significant fragmentation. Such a
collapse requires a massive gas cloud without significant rotation (which
would collapse to a rotationally supported disk rather than a massive black
hole), and without significant cooling from metals or molecular hydrogen, either of which would cause the cloud to fragment rather than collapse to a massive black hole.  
The gas content in primordial galaxies at sufficiently high redshift tends to
be pristine, thus lacking the metals needed for metal-line cooling, and a
nearby source of photo-dissociating Lyman-Werner radiation can disrupt and
prevent formation of molecular hydrogen \citep[see e.g. recent work
  by][]{Regan2017}. The net result is a massive cloud capable of collapsing
directly to a black hole seed (referred to as a `Direct Collapse Black Hole',
or DCBH) with mass $M_{\rm{BH,seed}} \sim 10^4-10^6 \: \mathrm{M_\odot}$
\citep[e.g.][]{HaehneltRees1993, LoebRasio1994, Begelman2006,
  ReganHaehnelt2009}. Moreover, high baryonic streaming
velocities with respect to the dark matter may promote the growth of DCBHs
\citep{Hirano2017, Schauer2017} with alternatively, or additionally
sufficiently high gas turbulent velocities preventing excessive fragmentation as well.  While DCBH seeds can be very massive (and
thus easier to grow to the high masses observed at $z \sim 6-7$), the strict
criteria for formation suggests they may be relatively rare to form (see
e.g. \citeauthor{Habouzit2016} \citeyear{Habouzit2016}, \citeauthor{Wise2019} \citeyear{Wise2019}
and recent reviews by \citeauthor{Latif2016} \citeyear{Latif2016} and
\citeauthor{Woods2019} \citeyear{Woods2019}).   

Given the uncertainty in formation mechanisms and resolution limitations,
cosmological simulations often rely upon a simple prescription to insert a
black hole seed of $\sim 10^5 \: \mathrm{M_\odot}$ into sufficiently well-resolved
haloes \citep[e.g.][]{Sijacki2009, DiMatteo2012, DeGrafBHGrowth2012,
 Dubois2014, Hirschmann2014, Nelson2015, Schaye2015, Feng2016}. This method
may be considered to broadly encompass each scenario described above: it is
comparable to a recently formed DCBH, or could be a PopIII or NSC remnant
which formed earlier and grew by several orders of magnitude before being
added to the simulation.  However the number densities of black hole and host
halo occupation fractions could differ largely. Furthermore, the final masses
of the largest black holes have been shown not to be strongly sensitive to the
seed mass provided it is rather large ($\sim 10^5 \mathrm{M_\odot}$ or above), and this
model 
has been shown to well reproduce observed populations of black holes
\citep[e.g.][]{Springel2005, DiMatteo2005, Sijacki2007, DeGrafBHGrowth2012,
  Hirschmann2014, Sijacki2015, DeGraf2015, Beckmann2017, DeGraf2017,
  DiMatteo2017, McAlpine2017, Weinberger2018}.  However, by construction these
models are unable to directly investigate early black hole formation and
growth. Furthermore, they treat each black hole's seed equivalently, and so
cannot address how sensitive individual black hole histories or time-dependent
statistical properties of the whole population are on the frequency or method
by which supermassive black hole seeds form. In particular, they typically
rely upon two hand-selected mass scales: the threshold halo mass above which
black hole seeds are inserted, and the mass of those seeds.  These simple
assumptions may not be realistic, and suggest that more advanced models for
seed formation should be explored, and there have been several recent analyses investigating this, using both simulations and semi-analytic models \citep[see, e.g.][]{Habouzit2017, Ricarte2018, Wang2019}. 
This is especially timely in view of the
flood of next generation observational data, from, for example, JWST, WFIRST,
eROSITA, Athena, LSST, IPTA, EPTA, NANOGrav, PPTA, and LISA which will push into the presently
inaccessible regime of lower
mass black holes and higher redshifts, together with entirely new constrains
from 
gravitational wave observations. Hence, in this paper we address these
questions by studying how changing the seed frequency and criteria for seed
formation can impact individual black hole growth and the overall populations
of black holes across cosmic time. 

To avoid the prohibitive computational expense of repeatedly re-running a
large cosmological simulation with many varying parameters for black hole
seeding, we developed a post-processing method of re-calculating the growth of
black holes, which we apply to the Illustris simulation \citep{Nelson2015}.
Using the full output of black holes and their surrounding gas properties at
every timestep, we are able to calculate the efficiency with which black holes
should grow, even if the frequency and conditions for seed formation are
changed.  We apply this method to a variety of prescriptions for seed
formation, including a less frequent stochastic seed model, a gas-spin model,
and a physically motivated model based on the mass, metallicity, and gas spin
of progenitor galaxies based on the DCBH mechanism.  For each of these models,
we study which galaxies we expect to form black holes, and how the growth
history of each individual black hole from the original simulation is changed
by each of these models.  Using the full set of re-computed black hole
histories, we characterize the impact on the black hole mass function and
quasar luminosity function, late-time accretion efficiencies, black hole-host
scaling relations, and black hole merger rates, quantifying how the frequency
and mass ratios of black hole mergers are impacted by each seed model.

The outline of the paper is as follows.  In Section~\ref{sec:method} we
discuss the Illustris simulation (Section~\ref{sec:simulation}) and
associated construction of galaxy merger trees (Section~\ref{sec:sublink}), our post-processing method of calculating black hole
evolution (Section~\ref{sec:growth}), and the seeding models used in our
analysis (Section~\ref{sec:seedingmodels}).  In Section~\ref{sec:results} we
discuss the results of our post-processing analysis.  Section~\ref{sec:seedproperties} shows how the seed model affects seed formation and
the galaxies in which seeds form; Section~\ref{sec:globalproperties} shows the
impact on global black hole statistics over cosmic time; and
Section~\ref{sec:mergers} shows the impact on black hole mergers. Finally, we
summarize our conclusions in Section~\ref{sec:conclusions}.

\section{Method}
\label{sec:method}
\subsection{Illustris Simulation}
\label{sec:simulation}

In this work we analyse the
Illustris\footnote{http://www.illustris-project.org; \citet{Nelson2015}.}
suite of simulations with a cosmological box $106.5 \: \rm{Mpc}$ on a side,
performed with 
the moving mesh code {\small AREPO} \citep{Springel2010}. We only consider the highest
resolution simulation incorporating the full physics model \citep[for further
  details see][]{Vogelsberger2014a, Genel2014, Sijacki2015} which has target
gas cell mass $m_{\rm gas} = 1.26 \times 10^6 \: \mathrm{M_\odot}$ and dark matter
particle mass $m_{\rm DM}=6.26 \times 10^6 \: \mathrm{M_\odot}$.  This run uses a standard $\Lambda$CDM cosmology, with $\Omega_{m,0}=0.2726$, $\Omega_{\Lambda,0}=0.7274$, $\Omega_{b,0}=0.0456$, $\sigma_8=0.809$, $n_s=0.963$, $H_0=70.4 \, \rm{km}\, \rm{s}^{-1} \, \rm{Mpc}^{-1}$ \citep[consistent with][]{Hinshaw2013}.

For the purpose of this study the modelling of black hole seeding, growth and
feedback is most relevant, so we briefly summarize it here. Black holes are
treated as collisionless sink particles. In Illustris, their seed mass is $10^5\: \mathrm{h^{-1} \: M_\odot}$ and they are seeded in all haloes with mass
above $5 \times 10^{10} \: \mathrm{h^{-1} \: M_\odot}$ if void of a black hole
particle. This seeding model is loosely motivated by the direct collapse scenario
\citep[see e.g.][]{HaehneltRees1993, Loeb1994, BrommLoeb2003, Begelman2006,
  ReganHaehnelt2009}, but may remain broadly consistent with lighter seed
formation models 
provided that their occupation fraction is similar and that they grow
efficiently enough in the early Universe. After seeding, black holes may grow
in mass either through gas accretion, which is modelled assuming a
Bondi-Hoyle-like rate \citep{BondiHoyle1944, Bondi1952} capped at the
Eddington limit, or by merging with other black holes which happen to be within
each others' smoothing lengths. In the absence of feedback such a black hole
growth prescription would result in unrealistically large black hole
masses. It is hence necessary to incorporate feedback prescriptions
which in Illustris is accomplished by invoking three separate modes:
``quasar'', ``radio'' and ``radiative'', which depend on the accretion
efficiency \citep[for more details, see][]{Vogelsberger2014b, Sijacki2015}.

We note that while this black hole model leads to a wide range of black hole
properties in good agreement with observational constraints
\citep{Sijacki2015}, the seeding prescription is rather simplistic. Thus, the
aim here is to explore different assumptions regarding seeding to pin down how
much they can affect the properties of a representative cosmological black hole
population, and in turn how this can inform us about the likely seeding
scenario.

\subsection{Constructing galaxy merger trees}
\label{sec:sublink}

The Illustris simulation identifies dark matter haloes using a
friends-of-friends (FOF) halo-finder \citep{Davis1985} with linking-length of
0.2 and baryonic matter assigned to the same group as the nearest dark
matter particle.  Following this, the SUBFIND algorithm \citep{Springel2001, Nelson2015} is run to produce a catalog of gravitationally self-bound substructures within each halo, which we use to determine galaxy properties.  Furthermore, to track galaxies over time (in particular to identify the progenitors of a given galaxy), we use the SubLink catalog produced by \citet{Rodriguez-Gomez2015}.  This catalog identifies descendant subhaloes across snapshots by summing a merit function (based on particle binding energies) for each particle common to a given pair of subhaloes in different snapshots.  The descendant is then defined as the subhalo with the highest summed merit score.  In this manner, we have access to a complete tree of all resolved subhaloes, such that we can identify galaxy mergers and the progenitors of any given galaxy, a key point when investigating if a given galaxy ever satisfied the conditions for SMBH formation at any earlier time.

\subsection{Post-processing black hole growth}
\label{sec:growth}

One of the challenges in investigating black hole seeding mechanisms with
cosmological simulations is the large computational expense to run such
simulations.  Ideally, these simulations must have high resolution to model
the regions in which black hole seeding occurs, large volume so as to include
rare objects such as massive high-redshift quasars \citep[see
  e.g.][]{Mortlock2011}, and be run to low-redshift to provide information on
the SMBH population in the local Universe where observational constraints are strongest (both for predictive purposes, and also to confirm the accuracy of the simulation).  Furthermore, the parameter space to investigate is large, spanning seeding environment (where and when black hole seeds form) and the seed black hole itself\footnote{Note that these considerations assume a fixed black hole accretion and feedback model, and that relaxing this would further enlarge the parameter space for exploration.}.  
Hence rather than re-running the entire Illustris simulation for each
variation in these parameters \citep[at a cost of $\sim$16 million CPU hours,
  see][]{Vogelsberger2014b}, which would be numerically prohibitive, we take a
post-processing approach in which we 
re-calculate the entire growth history of the full black hole population,
using the high temporal resolution output from the original run, which
contains local gas properties of each black hole at every timestep.

 Black hole accretion within the simulation is given by
\begin{equation}
\dot{M}_{\rm{BH}} = \frac{4 \pi \alpha G^2 M_{\rm{BH}}^2 \rho}{c_s^3}       
\label{eq:accretion}
\end{equation}
\citep[see][]{BondiHoyle1944, Bondi1952}, with an imposed upper limit of the Eddington rate [$\dot{M}_{\rm{Edd}\
}=(4 \pi G M_{\rm{BH}} m_p) / (\epsilon_r \sigma_T c)$].
Thus the accretion rate depends upon black hole mass ($M_{\rm{BH}}$) and the
density ($\rho$) and sound speed ($c_s$) of the nearby gas, with dimensionless
$\alpha = 100$, radiative efficiency $\epsilon_r = 0.2$, gravitational
constant $G$, speed of light $c$, proton mass $m_p$ and Thompson
cross-section $\sigma_T$.  Since the simulation saves each of these properties every time
the black hole properties are calculated (with typical timesteps on the order
of $2 \times 10^4$~yrs), all the information necessary to re-calculate
accretion rates is saved. 

We perform this growth recalculation by taking the necessary properties ($M_{\rm{BH}}$, $\rho$, $c_s$) to calculate the expected accretion rate (Equation~\ref{eq:accretion}, with imposed Eddington limit).  We repeat this calculation at each timestep to find the expected mass at the following timestep (i.e. $M_{\rm{BH},i+1} = M_{\rm{BH},i} + \dot{M}_{\rm{BH,i}} \Delta t$).
In this way we are able to reproduce the full accretion history of the entire black hole population, but with the freedom to change certain parameters (e.g. seeding a subset of the black hole population or changing the initial seed mass).
We also incorporate black hole mergers in our new calculation.  Because we are
using the gas properties local to black holes in the original simulation, our black holes are
necessarily subsets of the original Illustris black holes (though often with different
masses).  Thus the mergers for our new black hole populations are also
subsamples of the mergers occurring in the original simulation (and for each
merger, the simulation saves the merger time and the ID and mass of each black
hole).  We determine mergers in our recalculated history by extracting mergers in the original simulation which involve only black holes seeded in our new model, at which point we combine the masses of the merging black hole pair (using our newly-calculated mass histories).  

We note that black holes in the Illustris simulation merge as soon as they are
within the black hole smoothing length of one another, without considering the
hardening/merging timescale, which may delay when a given black hole pair
should merge.  In this analysis we use the merging times of the original
simulation, but note that an extended hardening times may delay black hole
mergers and thus impact our predicted black hole merger rate.  However, an initial investigation
finds that adding a delay of 1Gyr has minimal impact on the overall merger rates, except at the highest
redshifts where the first black hole mergers occur, consistent with results from \citet{Salcido2016}, who found minimal impact for delay timescales of 5Gyr in gas-poor galaxies. 

We note that the primary limitation of our model is the inability to
incorporate self-consistent feedback and self-regulation into post-processing
calculations.  In the original simulation, AGN feedback is deposited into the
surrounding gas as the black hole grows; in our post-processing analysis we
are able to modify the accretion history of the black hole, but the feedback
energy deposited remains unchanged.  At low black hole masses we expect this
to have a minimal impact, as the black holes are generally too small to have a
strong impact on the gas supply of their host halo.  At high black hole mass,
we know that quasar feedback plays an important role, and acts as a catalyst
for self-regulation of further growth \citep[e.g.][]{Sijacki2007,
  DeGrafBHGrowth2012}.  Thus our recalculated growth histories may
underestimate the final black hole mass, but this will only be relevant for black holes whose recalculated mass is substantially below the original mass at the time when feedback-induced self-regulation begins (we further discuss this issue and potential implications in Section \ref{sec:globalproperties}).  

We further note that the Illustris simulation output is incomplete for black holes below $z \sim 0.38$. This missing data necessarily prevents recalculating growth and mergers at lower redshifts, so we limit our analysis to $z \geq 0.5$.  This limitation has a minimal impact on our results, however, as it is after the majority of black hole growth.  We do not expect any qualitative changes at lower redshifts.
\begin{figure}
    \centering
    \includegraphics[width=8.5cm]{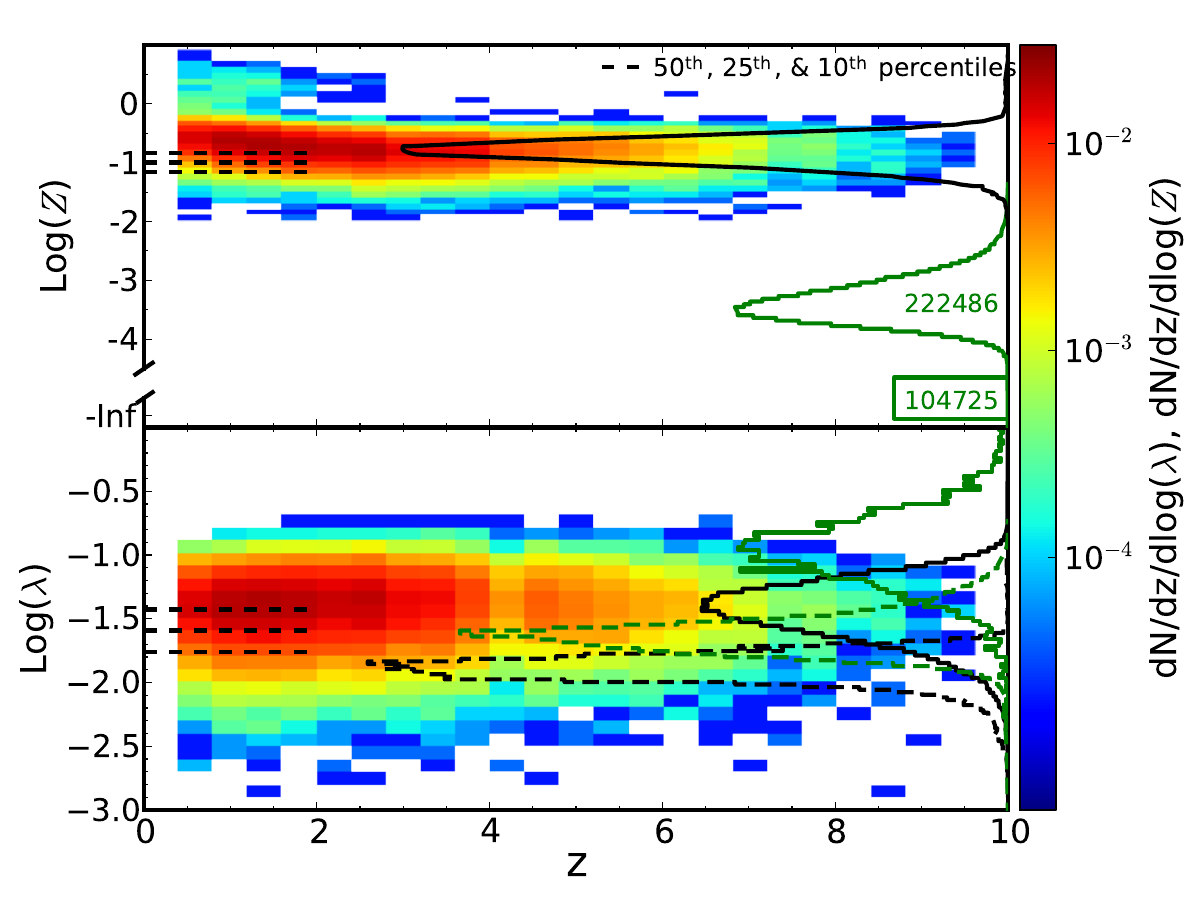}
    \caption{\textit{Top:} Colourmap showing the distribution in
      metallicity-redshift space of newly seeded haloes in the original
      Illustris simulation, with dashed horizontal lines showing the 50th,
      25th, and 10th percentiles.  The solid black histogram shows the 1D
      distribution of metallicity for those galaxies, while the solid green
      histogram shows the equivalent distribution for z = 4 galaxies above our
      mass threshold for DCBH seeding ($3 \times 10^9 \: \mathrm{M_\odot}$,
      see Section~\ref{sec:DCBH_seeding}).  Note that a large fraction of
      galaxies have metallicity $Z = 0$, which is depicted as a single block
      at the bottom of the panel, despite the logarithmic scaling. Numbers within the 1D histograms specify the number of galaxies from the simulation.
\textit{Bottom:} Same as the top panel, but for gas spin $\lambda$. In
addition to the solid histograms showing the distribution of $\lambda$ for
newly-seeded galaxies in the original simulation (black) and for z = 4
galaxies with $M > 3 \times 10^9 \: \mathrm{M_\odot}$ (green), we also show
the equivalent distributions for $\lambda_{\rm{max}}$, the maximum spin for
DCBHs to form (dashed lines; see Section~\ref{sec:DCBH_seeding}).   
}
\label{fig:spin_vs_z}
\end{figure}

\subsection{Seeding models}
\label{sec:seedingmodels}

\subsubsection{Illustris seeding}
\label{sec:illustris_seeding}

Within our framework, the simplest seeding prescription is to re-use the data from seeding black holes in the original Illustris simulation, i.e. when a halo crosses a specified threshold of $5 \times 10^{10} \: \mathrm{h^{-1} \,M_\odot}$.  By only seeding black holes that were originally seeded in Illustris, we explicitly have the local gas properties necessary for re-calculating growth, and the black hole mergers in this new case will simply match the black hole merger trees from Illustris.  If using the same seed mass as Illustris ($10^5 \: \mathrm{h^{-1} \,M_\odot}$), this will mean that each black hole should be `re-grown' in a way that matches the data from the original simulation, which we have explicitly confirmed. 

\subsubsection{Stochastic seeding}
\label{sec:random_seeding}

An obvious variation on using the Illustris seed criterion is to simply seed a
fixed fraction ($f_{\rm{seed}}$) of Illustris-seeded black holes, accomplished
using a random number generator to determine whether a given Illustris seed
should be considered `seedable' in our re-calculation.  As this model
continues to only seed subset of black holes which have been seeded by
Illustris, we again have the complete set of local gas properties.
Furthermore, since all black holes are chosen from the original Illustris
catalog, all black hole mergers can be extracted from the original Illustris
merger trees.

\begin{figure}
    \centering
    \includegraphics[width=8.5cm]{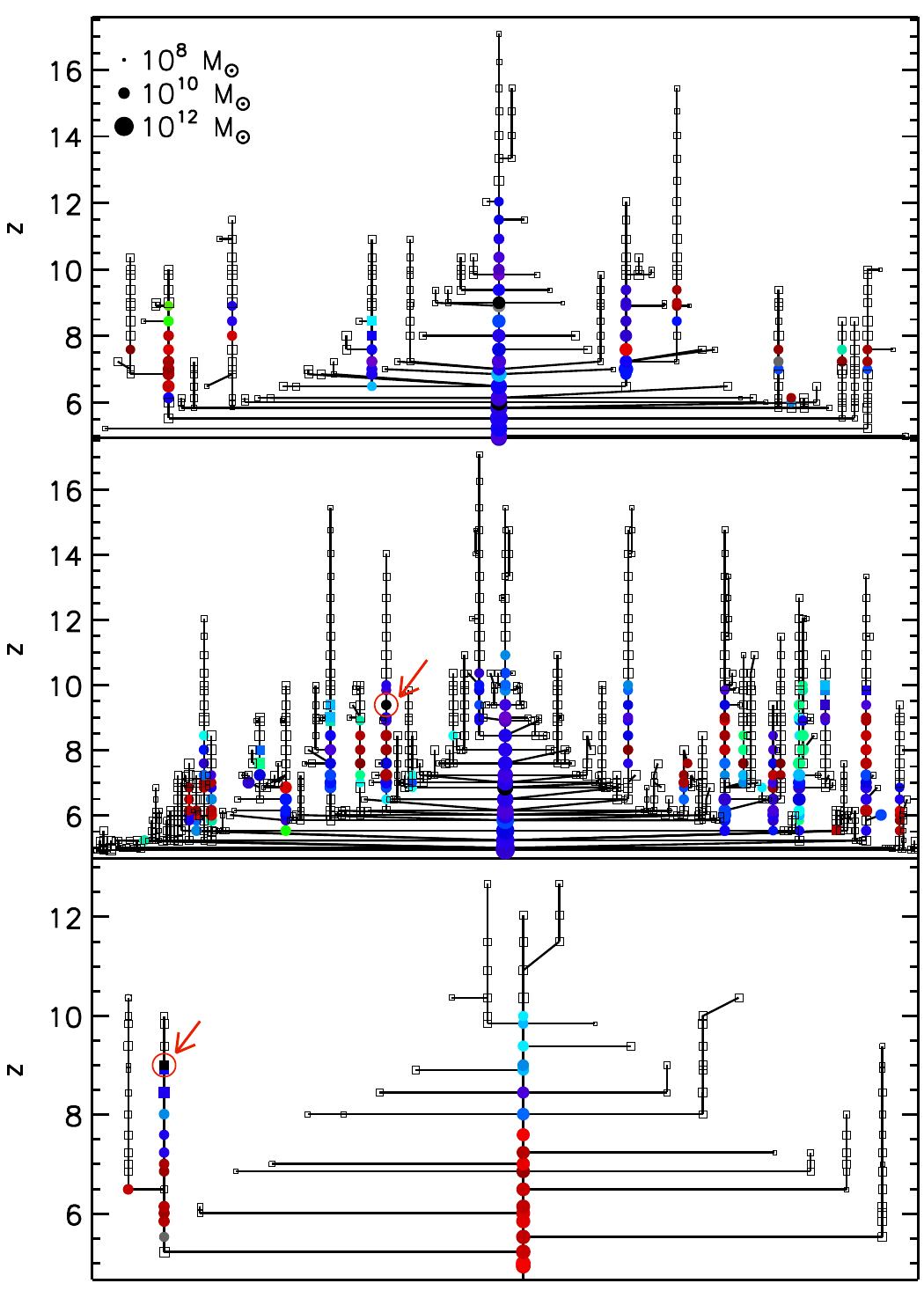}
    \caption{Three sample galaxy progenitor trees. \textit{Top:} a
      moderate-size progenitor tree, with no seedable progenitors.
      \textit{Middle:} A very large progenitor tree. \textit{Bottom:} A
      typical example of a seedable progenitor tree.  Black open symbols mean
      no gas (and thus no gas spin); filled symbols are colour-coded by gas
      spin.  Circles are galaxies with non-zero stellar mass, squares are
      galaxies with zero stellar mass.  Galaxies satisfying our criteria for
      DCBH formation (see Section~\ref{sec:DCBH_seeding}) are circled in red
      and indicated by red arrows.}
\label{fig:progenitor_galaxies}
\end{figure}

We note that although we seed using a fixed probability, the total number of black holes is not the same fraction of original black holes (i.e. $N_{\rm{BH,f_{seed}}} \ne f_{\rm{seed}} \times N_{\rm{BH,Illustris}}$), due to black holes which involve multiple black hole seeds that eventually merge together.  Furthermore, this is a mass-dependent effect, since high-$M_{\rm{BH,Illustris}}$ black holes tend to have undergone the largest number of mergers and thus have the largest number of chances to have at least one seed in our new calculation (e.g. for $f_{\rm{seed}}=0.1$, at z = 0.5 $18\%$ of Illustris black holes above $10^6 \: \mathrm{M_\odot}$ have been reseeded in our new model, compared to $58\%$ of Illustris black holes above $10^8 \: \mathrm{M_\odot}$. This is discussed further in Section~\ref{sec:globalproperties} and especially Figure~\ref{fig:nprogs}). This different impact on high- and low-mass black holes means that the impact on overall black hole populations, such as the black hole mass function and luminosity function as functions of cosmic time, is more complicated than a simple decrease in total black hole number (for further details, see Section~\ref{sec:globalproperties}).

\subsubsection{Spin- and metallicity-based seeding}
\label{sec:spin_seeding}

In Figure~\ref{fig:spin_vs_z} we investigate the galaxies seeded by black holes
in the Illustris simulation, showing the distribution in both spin-redshift
space and 
metallicity-redshift space (both important criteria in direct-collapse
formation models), where $\lambda$ and $Z$ are computed for the gas
within the gas half-mass radius of the galaxy.  This distribution shows that
gas spin in the center of newly black hole-seeded galaxies is essentially
redshift independent, and metallicity only shows a slight increase at low
redshift (below $z \sim 1.5$).  However, this is explicitly the case for
haloes which have just recently crossed the $5 \times 10^{10} \:
\mathrm{h^{-1} \, M_\odot}$ threshold; we consider the impact on other galaxies in more detail in Section~\ref{sec:seedproperties}. 

As a variant on the stochastic seeding model, here we seed a subset of
Illustris black holes based on the dimensionless spin parameter of gas within
the gas half-mass radius of the host galaxy at the time when the Illustris
black hole is seeded \citep[as low-spin may be needed for gas to collapse
  directly to a black hole; see Section~\ref{sec:DCBH_seeding} and][for more
  details]{LodatoNatarajan2006, NatarajanVolonteri2012}.  Specifically, we
impose a cut on spin such that only black holes in galaxies with $\lambda <
\lambda_{\rm{crit}}$ are seeded, where we select $\lambda_{\rm{crit}}$ to
correspond to a given fraction of galaxies (i.e. to be comparable to a given
value of $f_{\rm{seed}}$).  Since $\lambda$ is approximately
redshift-independent, seeding based on gas spin in this manner is
qualitatively similar to the random seed selection model (note that the 50th-,
25th-, and 10th-percentiles are marked in Figure~\ref{fig:spin_vs_z},
corresponding to $f_{\rm{seed}} = 0.5, 0.25, 0.1$ in Section~\ref{sec:random_seeding}).  As a further test, we imposed an additional criterion based on gas metallicity (see also Section~\ref{sec:DCBH_seeding}).  The metallicity distribution is also roughly redshift-independent for $z > 2$, so this additional criterion tends to only affect a small number of low-redshift, low-mass black holes, and otherwise produces qualitatively similar results.

\begin{figure}
    \centering
    \includegraphics[width=8.0cm]{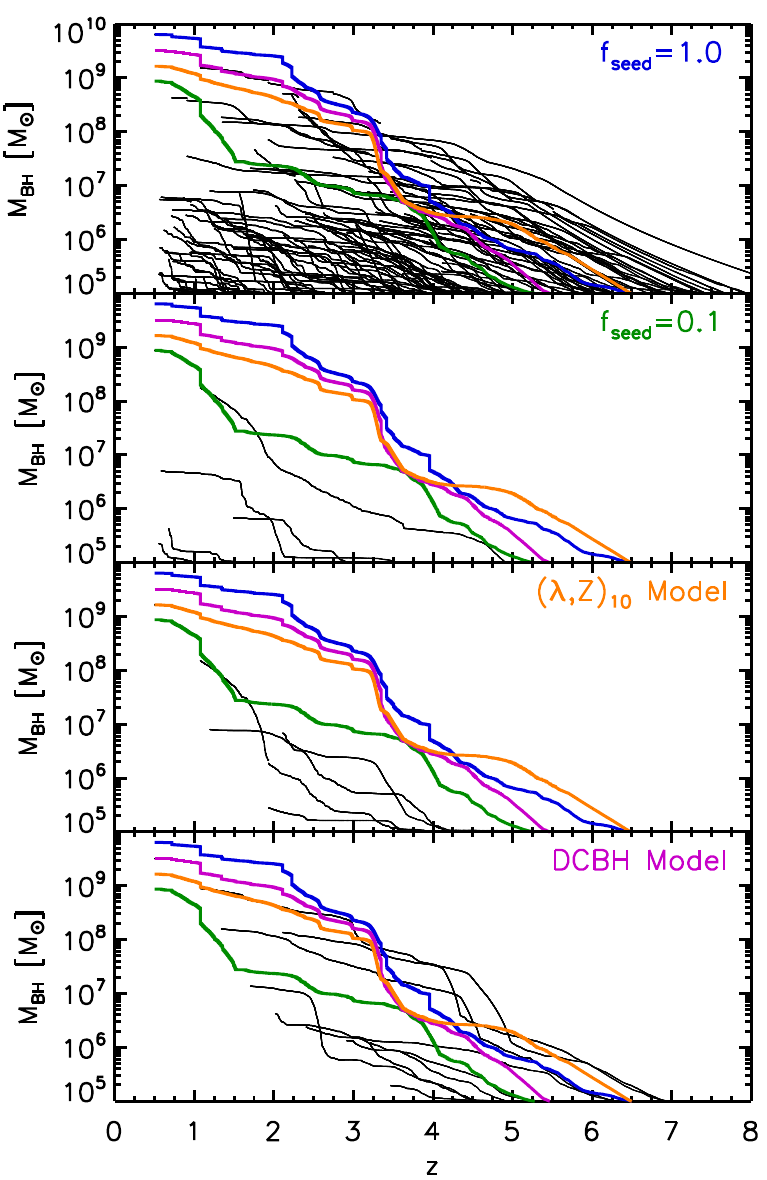}
    \caption{Growth history of an entire black hole merger tree using the original
      Illustris seeding ($f_{\rm{seed}}=1$), using $f_{\rm{seed}}=0.1$,
      seeding 10\% of black holes based on the lowest instantaneous spin and
      metallicity when Illustris seeds them ($(\lambda,Z)_{10}$ model) and
      using the DCBH seeding model. The coloured lines are the same on all
      panels, showing the most massive progenitor history for these four
      seeding models. Both the final mass of the most massive black hole and
      the number of black holes that merge with it is significantly affected
      by the seeding models explored.}
    \label{fig:progenitor_histories}
\end{figure}

\subsubsection{Progenitor-based direct collapse seeding}
\label{sec:DCBH_seeding}

The models described in Sections~\ref{sec:random_seeding} and
\ref{sec:spin_seeding} are based solely upon instantaneous properties when
a black hole particle was first inserted into the Illustris simulation.
However, we note that the insertion of a sink particle in Illustris (as in
other cosmological simulations) is not necessarily intended to represent the
actual formation of a seed black hole, but rather a point at which we would expect a seed to have previously formed and grown to a ``resolvable'' mass.
A more physically meaningful approach is to consider which black holes seeded
in Illustris have a progenitor galaxy which satisfies the necessary criteria
for the formation of a black hole seed, in this case by direct collapse.  We
are able to do this by considering the full history of all Illustris galaxies
(and the merger trees involving them; see Section~\ref{sec:sublink}), checking
for a set of conditions necessary for DCBHs where we impose three criteria:

\begin{itemize}
  \item Galaxy mass: We impose that $M_{\rm{subhalo}} > 3 \times 10^9 \: \mathrm{M_\odot}$ (where $M_{\rm{subhalo}}$ is the total mass contained within the SUBFIND-defined subhalo, as discussed in Section~\ref{sec:sublink}).  This is largely a resolution-based criterion (corresponding to $\sim 500$ dark matter particles) such that we have sufficient gas particles to reasonably resolve the remaining criteria.
\end{itemize}

\begin{itemize}
  \item Gas spin: Sufficiently high gas spin implies a more gravitationally
    stable structure which may not lead to the further collapse necessary to form a DCBH seed. A rough approximation for this criterion is
\begin{equation}
\lambda < \frac{m_d^2 \,Q_c\, T_{\rm vir}}{8 \,j_d \,T_{\rm{gas}}},
\label{eq:spin}
\end{equation}
where $m_d$ is the fraction of gas involved in infall, $Q_c$ is the Toomre stability parameter, $j_d$ is the angular momentum in the disc, $T_{\rm{vir}}$ is the virial temperature of the halo, and $T_{\rm{gas}}$ is the gas temperature \citep{LodatoNatarajan2006}. We calculate $T_{\rm{vir}}$ and $T_{\rm{gas}}$ directly from the simulation, noting that $T_{\rm{gas}}$ is typically on the order of $1-3 \times 10^4 $K.  The remaining parameters ($m_d, j_d$, and $Q_c$) are much more sensitive to the structure of the disk which is less well-resolved in our simulation, especially in the low-mass galaxies at high-$z$ relevant for the DCBH formation.  As such, we follow \citet{NatarajanVolonteri2012} and use $m_d = j_d = 0.05$ and $Q_c = 2$ rather than use the resolution-sensitive values from the simulation.

We show an example of these constraints using 1D histograms in
Figure~\ref{fig:spin_vs_z}.  The solid (dashed) black histogram shows the
distribution of $\lambda$ ($\lambda_{\rm{max}}$) of galaxies newly-seeded in
the original Illustris simulation, where there is a minor overlap (i.e. $\lambda < \lambda_{\rm{max}}$).  For comparison, the green histograms show the distribution for $\lambda$ and $\lambda_{\rm{max}}$ for all galaxies above $3 \times 10^9 \mathrm{M_\odot}$ (our resolution-based mass threshold for DCBH seeding) at $z = 4$.  Here we see that both $\lambda$ and $\lambda_{\rm{max}}$ are higher than for the newly-seeded haloes in Illustris, but again there is a region of overlap where in principal DCBH formation can occur.

  \item Gas metallicity: In addition to rotational support, one of the main
    obstacles for massive DCBH seed formation is fragmentation of the
    collapsing gas cloud.  Specifically, metals within the gas can facilitate
    cooling, which leads to fragmentation of the cloud prior to collapse into
    a single supermassive black hole seed.  To account for this, we impose a
    metallicity threshold of $Z < 10^{-5} Z_\odot$ for the galactic gas
    \citep[consistent with][]{DevecchiVolonteri2009}.  We note that our
    results are not sensitive to the exact threshold, since Illustris galaxies
    rarely have non-zero metallicity near this value (e.g., at $z > 5$,
    $<0.05\%$ of galaxies with non-zero metallicity have metallicity below
    $10^{-3} Z_\odot$). This is also shown in Figure~\ref{fig:spin_vs_z},
    where the green 1D histogram shows the distribution of metallicities for
    galaxies above $3 \times 10^9 \: \mathrm{M_\odot}$ at $z = 4$: the peak of
    non-zero metallicity galaxies occurs at $Z \sim 10^{-3.5} \: Z_\odot$, and
    nearly one-third of galaxies have zero metallicity. 

\end{itemize}

We note
      however that this is sensitive to the metallicity enrichment model of
      the simulation. In the Illustris simulation, star particles are treated
      as single-age stellar populations which deposit metals into the
      neighboring gas cells based on the stellar evolution model, which
      includes supernovae (including core collapse, see
      \citeauthor{Portinari1998} \citeyear{Portinari1998} and
      \citeauthor{WoosleyWeaver1995} \citeyear{WoosleyWeaver1995}, and Type Ia, see \citeauthor{Thielemann2003} \citeyear{Thielemann2003}) and AGB winds \citep[see][]{Karakas2010} [note that the simulation directly models nine distinct elements, though we only consider total metallicity in this analysis].  The metals are also distributed by stellar feedback via metal-loaded winds, and metal mixing is modeled purely by advection \citep[for complete details of the models used, see][]{Vogelsberger2013}.  

Our results may be sensitive to this model as a less efficient metal
mixing model could lead to more haloes satisfying the metallicity criteria;
conversely, a full incorporation of PopIII star formation may contribute more
high-redshift metallicity, thus decreasing the haloes satisfying our metal
criterion as there is no high redshift metallicity floor in Illustris.

These criteria are used to determine which Illustris black holes have progenitor galaxies with the necessary conditions for DCBH seed formation.  Specifically, for the host subhalo of every newly seeded black hole in the Illustris simulation we check every progenitor galaxy in the SubLink galaxy merger tree \citep[see Section~\ref{sec:sublink} and][]{Rodriguez-Gomez2015}.  Any Illustris black hole which has at least one progenitor galaxy simultaneously satisfying all three criteria is considered a `seedable' black hole according to our DCBH model. We note here that our method of post-processing black hole growth relies on the output data provided for each Illustris black hole. As such, we are unable to directly calculate growth for black holes which do not exist in the original simulation. Thus our seeding model here does not seed black holes as soon as a galaxy satisfies the DCBH criteria, but rather when the original simulation seeds a black hole which had a progenitor galaxy which satisfies the criteria. 

\begin{figure*}
    \centering
    \includegraphics[width=15.0cm]{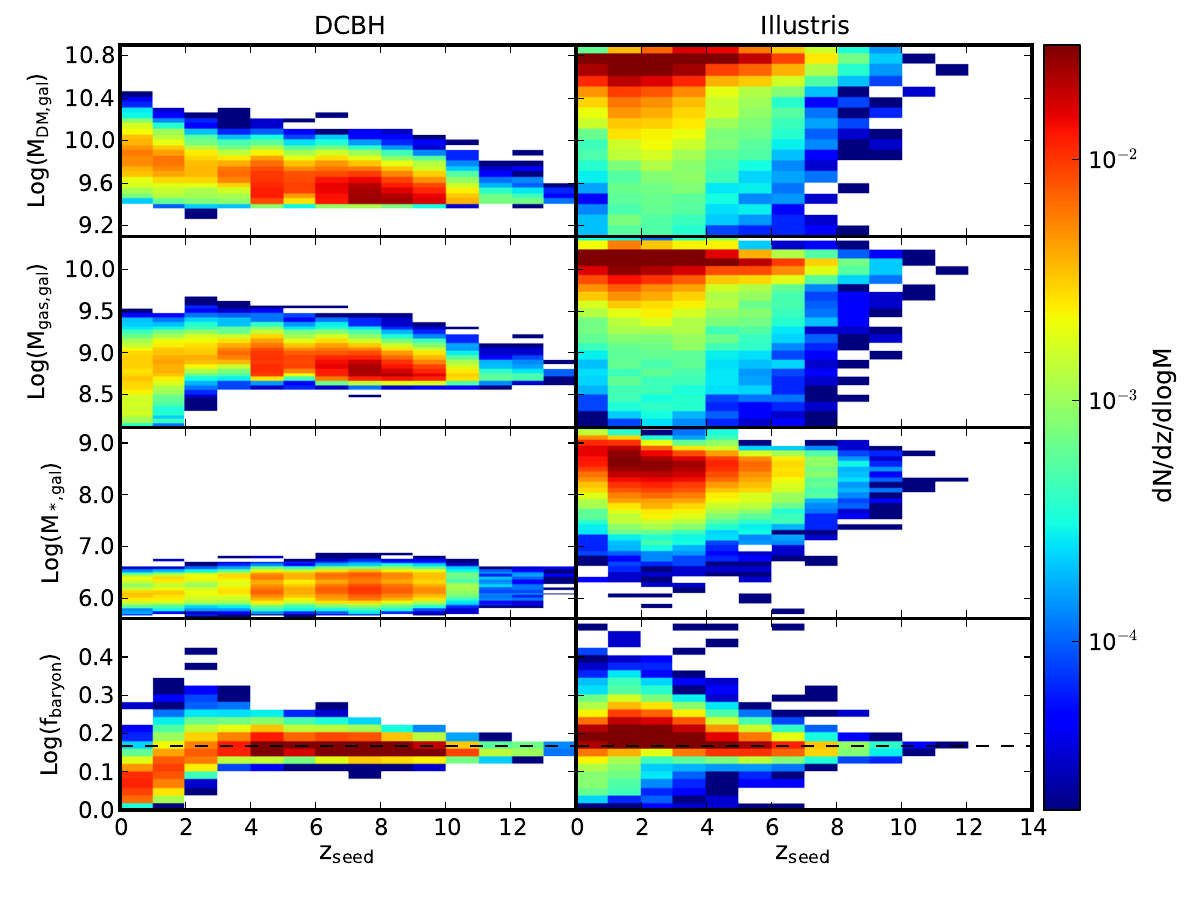}
    \caption{Distribution of host galaxy (based on subhalo, rather than overall halo) dark matter mass ($M_{\rm{DM}}$), gas
      mass ($M_{\rm{gas}}$), stellar mass ($M_*$), and baryon fraction
      ($f_{\rm{baryon}}$) for newly-seedable galaxies in the DCBH model (left)
      and the original Illustris simulation (right).  In the $f_{\rm{baryon}}$
      panels, the universal baryon fraction is shown as a dashed black
      line. The fixed $M_{\rm{halo}}$ threshold used for seeding black holes
      in Illustris results in very little redshift evolution, and many seeds
      forming at late times.  In contrast, the DCBH model seed primarily at
      high-redshift, in lower-mass galaxies, and with more substantial
      evolution in dark matter and gas masses at low redshift.}
    \label{fig:host_mass_vs_z}
\end{figure*}

To illustrate our DCBH seeding method, in Figure~\ref{fig:progenitor_galaxies}
we show the full galaxy progenitor trees of three $z = 5$ galaxies with
newly-seeded black holes in Illustris, colour-coded by gas spin and symbols
distinguishing between galaxies with $M_* = 0$ and $M_* > 0$ (vertical lines
represent individual galaxy evolution with time; horizontal lines represent
galaxy mergers).  The first is a representative moderate-sized galaxy tree
with no seedable progenitors; the second is a very large merger tree with a
single seedable progenitor galaxy (circled in red), and the third shows a
representative small merger tree containing a single seedable progenitor
galaxy (circled in red).

As seen in Figure \ref{fig:progenitor_galaxies}, the conditions for DCBH seed formation typically occur significantly earlier than the Illustris seeding.  As such, although we apply our post-processing analysis to a given black hole with an initial seed mass of $10^5 \rm{M}_{\odot}$ starting when Illustris seeds, there is a substantial period of time since actual seeding (when DCBH conditions are satisfied) during which the seed could grow. For high-redshift seeds, black holes tend to have several hundred Myr in which to grow, introducing two additional parameters to characterize this early growth: the seed mass when the initial DCBH forms, and the typical growth rate from that time until Illustris seeds the BH.  Among $z > 5$ seeds, we find that if the initial seed mass is $10^4 \rm{M}_{\odot}$, 96\% of DCBHs would be able to reach $10^5 \rm{M}_{\odot}$ by the time they are included in Illustris if limited to sub-Eddington rates, and 83\% would be able to do so even if initially formed at only $10^3 \rm{M}_{\odot}$.  As such, we find that the seed mass used in Illustris is fully consistent with a wide range of lower formation masses, followed by a reasonable period of sub-Eddington accretion. Recent work by \citet{Wang2019} investigated this early growth directly, using {\small GADGET-3} simulations which seed at $M_{\rm{BH,seed}}=10^3 \: \mathrm{M_\odot}$ based on local gas properties, and found that the mean mass for black holes in haloes at the threshold for seeding in Illustris is $\sim 10^{5.26} \: h^{-1} \: \mathrm{M_\odot}$, confirming that a seed mass of $\sim 10^5 \: \mathrm{M_\odot}$ is reasonable, and that we can expect our DCBH seeds to reach this mass by the time the Illustris black holes are included.

We further note that an additional constraint for DCBH formation is a nearby
source of photo-dissociating Lyman Werner radiation, capable of dissociating
molecular hydrogen and thus prevent fragmentation resulting from molecular
cooling.  However, the Lyman-Werner intensity threshold needed remains very
poorly constrained \citep[see, e.g.][who find $J_{\rm{crit}}$ ranges from $5
  \times 10^{-22}$ to $10^{-18}$ erg s$^{-1}$ cm$^{-2}$ sr$^{-1}$
  Hz$^{-1}$]{Agarwal2016}. Furthermore, recently \citet{Wise2019} proposed
that lower Lyman-Werner fluxes can still be sufficient, with the primary
driver for DCBH formation being high gas inflow rates onto the halo.  Hence
estimating for the incoming Lyman-Werner flux in the Illustris simulation would
be highly uncertain, given the resolution limitations and the lack of
PopIII stars.  Given these caveats, we do not include any
Lyman-Werner constraint in our analysis here.  Note that we expect our seed
model to over-estimate the DCBH formation rate compared to a simulation which
fully incorporates this additional criterion, and thus may be considered an
upper limit on the formation efficiency.

This DCBH seeding model is the most physically meaningful of all the models we consider, since we only take into account black holes in galaxies which have, at some point in the past, satisfied the necessary criteria for DCBH formation.  Because our post-processing growth calculations rely upon the local gas properties which are only saved around Illustris black holes, this model by construction only includes black holes which are also seeded in the original Illustris simulation.  As such, we again have the complete history of any mergers involving any combination of black holes seeded according to this new model.  Note that this approach ignores galaxies which satisfy the criteria for DCBH seed formation but which never have a black hole form in Illustris.  However, the only cases in which this can occur are galaxies which satisfy the above criteria but 
whose host haloes do not cross the $5 \times 10^{10} \: \mathrm{h^{-1} \,M_\odot}$ mass threshold for seeding.  These black holes would likely not grow into the regime in which we consider our results well-resolved ($M_{\rm{BH}} > 10^6 \: \mathrm{M_\odot}$), and so will not largely impact our results.  

\begin{figure*}
    \centering
    \includegraphics[width=8.5cm]{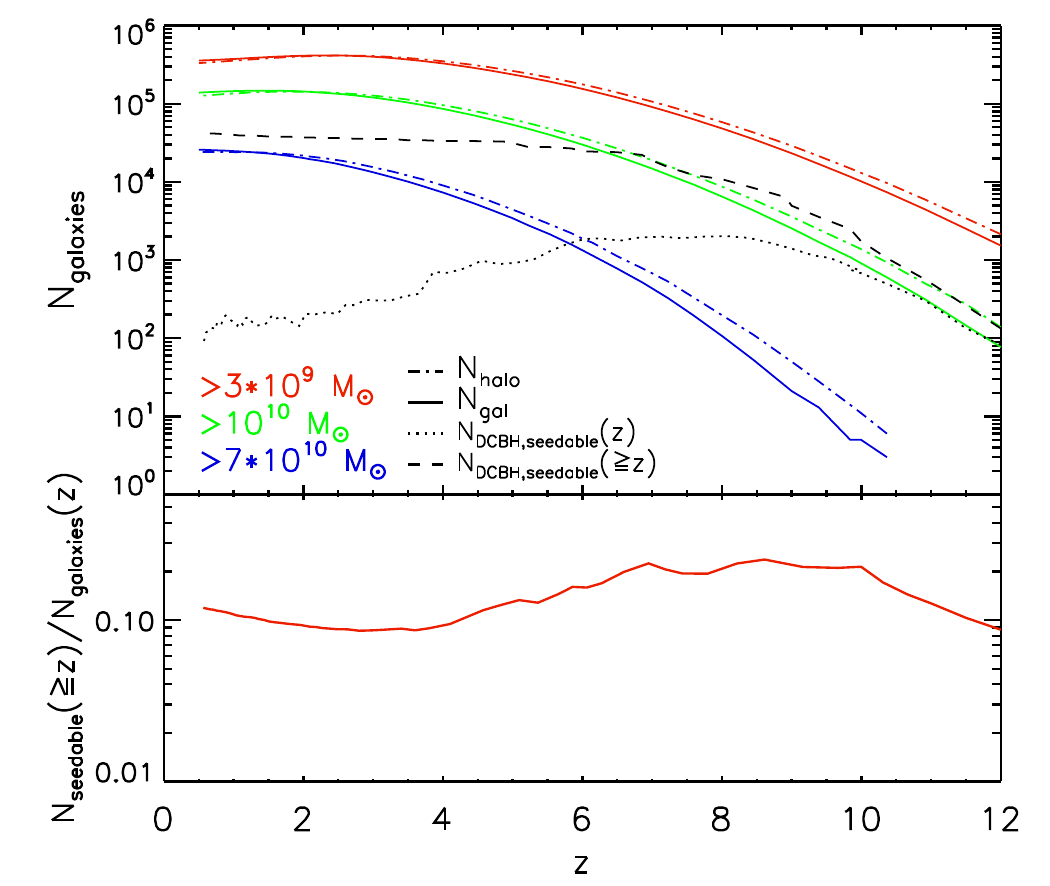} 
    \includegraphics[width=8.9cm]{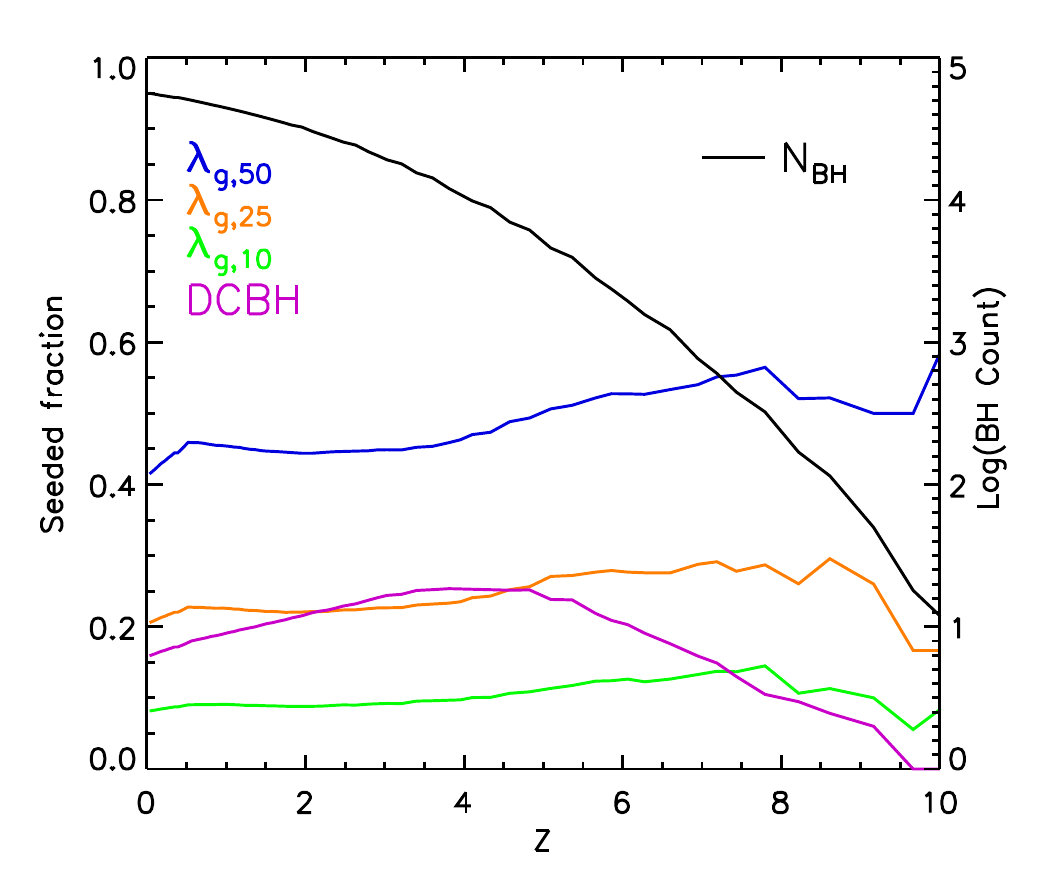}  
   \caption{\textit{Top Left:} Number of galaxies (solid lines) and haloes (dot-dashed lines) at the given redshift above a
      given mass cut, compared to the number of galaxies at
      redshift $z$ satisfying criteria for seeding in the DCBH model (dotted
      line) and the cumulative number of black holes seeded according to the
      DCBH seeding model (dashed line). Note that a halo mass of $\sim 7 \times
      10^{10} \: \mathrm{M_\odot}$ is the threshold for black hole seeding in
      the original Illustris simulation, so the blue line roughly corresponds
      to the number of black holes in Illustris.  \textit{Bottom left:} Ratio
      between the cumulative number of seedable galaxies with redshift $\ge z$
      and the total number of galaxies at redshift $z$ with $M_{\rm{galaxies}}
      \ge 3 \times 10^9 \: \mathrm{M_\odot}$ (the minimum mass considered for
      seedability, see Section~\ref{sec:seedingmodels}). \textit{Right:}
      Fraction of black holes from original Illustris simulation which are
      seeded if seeding is determined explicitly by gas spin, and our full
      DCBH seeding model.  The black curve also shows the total
      number of black holes seeded in Illustris as a function of redshift.} 
    \label{fig:seednumbers} 
\end{figure*}

We note that we are only able to check progenitor galaxy conditions in the saved snapshots, which sets the time-resolution for this model. This timescale ($\sim 40$ Myr for $z \sim 5-10$) is longer than the typical timescale for DCBH formation, such that it is possible to miss a galaxy which satisfies the criteria during a short time period between snapshots.  
The typical variation of $\lambda$ is relatively slow compared to the snapshot timescales (see the generally gradual colour variation in individual vertical lines in Figure~\ref{fig:progenitor_galaxies}), so we would expect this to be a sub-dominant effect.  The metallicity evolution can be fairly rapid, however, suggesting more frequent snapshots might show additional galaxies which satisfy the conditions for DCBH formation.  By sub-sampling the available snapshots and extrapolating to shorter inter-snapshot timescales, if we assume that the collapse time is on the order of $\sim 5$ Myr, it is possible we might underestimate the seed frequency by up to 50\%.  We expect this is a conservative upper limit based on significant extrapolation (and the unresolved metal enrichment); nonetheless, we would expect that even an increase at this upper limit should not qualitatively change the results presented here.  A more precise treatment would require a full re-simulation which checks for seed conditions on-the-fly, which is beyond the scope of this analysis.

\subsubsection{Individual black hole growth}
\label{sec:sample_growth}

In Figure~\ref{fig:progenitor_histories} we illustrate our post-processing
technique, showing the full history of a SMBH and all its progenitors using
the four different seeding models described above. Each panel shows the mass
history for the most massive black hole from the original Illustris simulation
as well as all black holes it merged with, having re-calculated the history
according to each seed model.  Over-plotted in colour on each panel is the most
massive progenitor history for each of the four models, determined by tracking
the final black hole back in time, and at each black hole merger selecting the
more massive progenitor black hole at the time of the merger (and as such the
coloured lines are identical in all four panels). There are significant
differences in the total number of seeded black holes and the masses such
black holes reach at both early and late times.

The top panel shows the history for the $f_{\rm{seed}}=1$ model, i.e. the
seeding conditions as in the original Illustris simulation.  This seed model
involves a 
large number of black hole mergers, with black holes being seeded continually throughout cosmic time.  The second panel shows the $f_{\rm{seed}}=0.1$ model.  By construction this model exhibits an equivalent seeding pattern as the original simulation, merely more rare.  This shows that forming only a subset of the original seed population can have a significant impact on the overall black hole growth history, affecting not only the seed number but also how efficiently the most massive progenitor grows (comparing the dark green and blue curves), and even the final mass (in this case resulting in a black hole $\sim$0.5 dex less massive).  In the third panel, we show seeding using a cut on the instantaneous gas spin and metallicity when Illustris first seeds black holes, as described in Section~\ref{sec:spin_seeding}.  Again we see a moderate impact on the final black hole mass, as well as a significant impact on the growth histories reaching that point.

Finally, in the bottom panel of Figure~\ref{fig:progenitor_histories} we show
the analogous history according to the DCBH seeding model.  Again we see a
moderate impact on the final $M_{\rm{BH}}$ (though weaker than seeding with
$\sim 10\%$ efficiency). We furthermore note a dramatic impact on when DCBH
seeds form: rather than forming continuously throughout the simulation, the
seeding occurs preferentially at high redshifts ($z > 4$) when gas clouds are
more likely to consist of pristine gas capable of collapsing directly to a
massive black hole without fragmenting.  This example growth history
demonstrates the impact the seed formation can have on black hole populations,
which we investigate quantitatively throughout the remainder of the paper.

\section{Results}
\label{sec:results}

\subsection{Black hole seeds and their host galaxies}
\label{sec:seedproperties}

Having performed the full recalculated growth histories of black holes
according to the models described in Section~\ref{sec:seedingmodels}, we
consider the galaxies in which seeds form in each model.  In
Figure~\ref{fig:host_mass_vs_z} we show the distribution of dark matter mass
($M_{\rm{DM}}$), gas mass ($M_{\rm{gas}}$), stellar mass ($M_*$), and baryon
fraction ($f_{\rm{baryon}}$) of galaxies with newly-seeded black holes as a
function of redshift, for the original Illustris seeding model (right panels)
and our DCBH-based seeding (left panels). The most obvious differences between
the seeding models is that the original Illustris seeding tends to take place
in galaxies with higher masses and at lower redshifts than our DCBH-based
seeding.  This is to be expected, since the idea behind Illustris seeding (and
that of most cosmological simulations) is to insert a black hole roughly
consistent with the black hole - host galaxy scaling relation possibly well
after the actual seed should have formed.  In contrast, our new DCBH model
finds galaxies in which the conditions for DCBH formation are satisfied, which
tend to occur at earlier times and in lower-mass galaxies \citep[similar to previous analyses, e.g.][]{Habouzit2017}. 

\begin{figure}
    \centering
    \includegraphics[width=8.5cm]{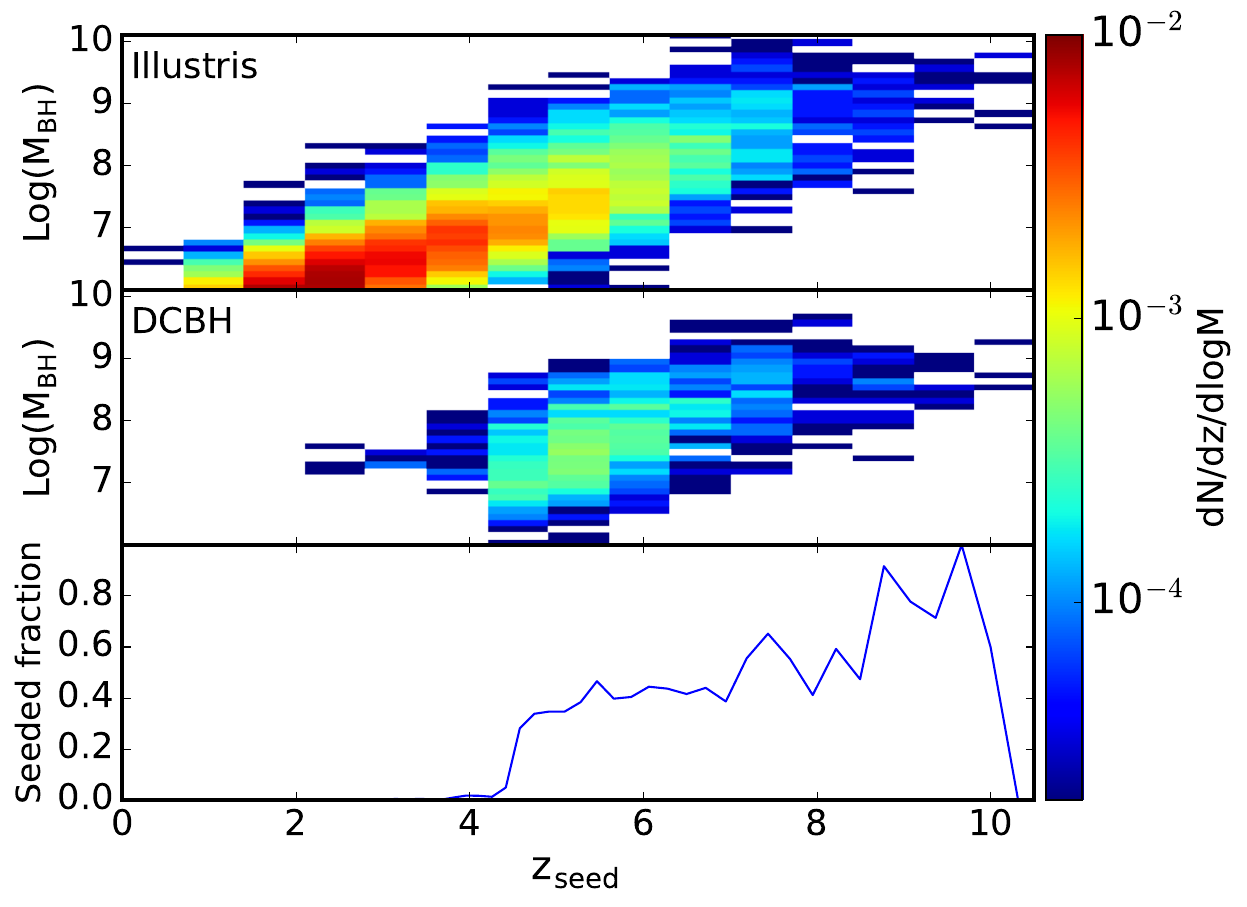}
    \caption{Distribution of final black hole mass, $M_{\rm{BH}} (z=0.5)$, as
      a function of redshift at which the black hole is first seeded,
      $z_{\rm{seed}}$.  \textit{Top:} Original Illustris data.
      \textit{Middle:} Our new DCBH seed model.  \textit{Bottom:} Fraction of
      Illustris-seeded black holes which are seeded in the DCBH seed model.}
    \label{fig:finalmass_vs_z}
\end{figure}

\begin{figure*}
    \centering
    \includegraphics[width=15.0cm]{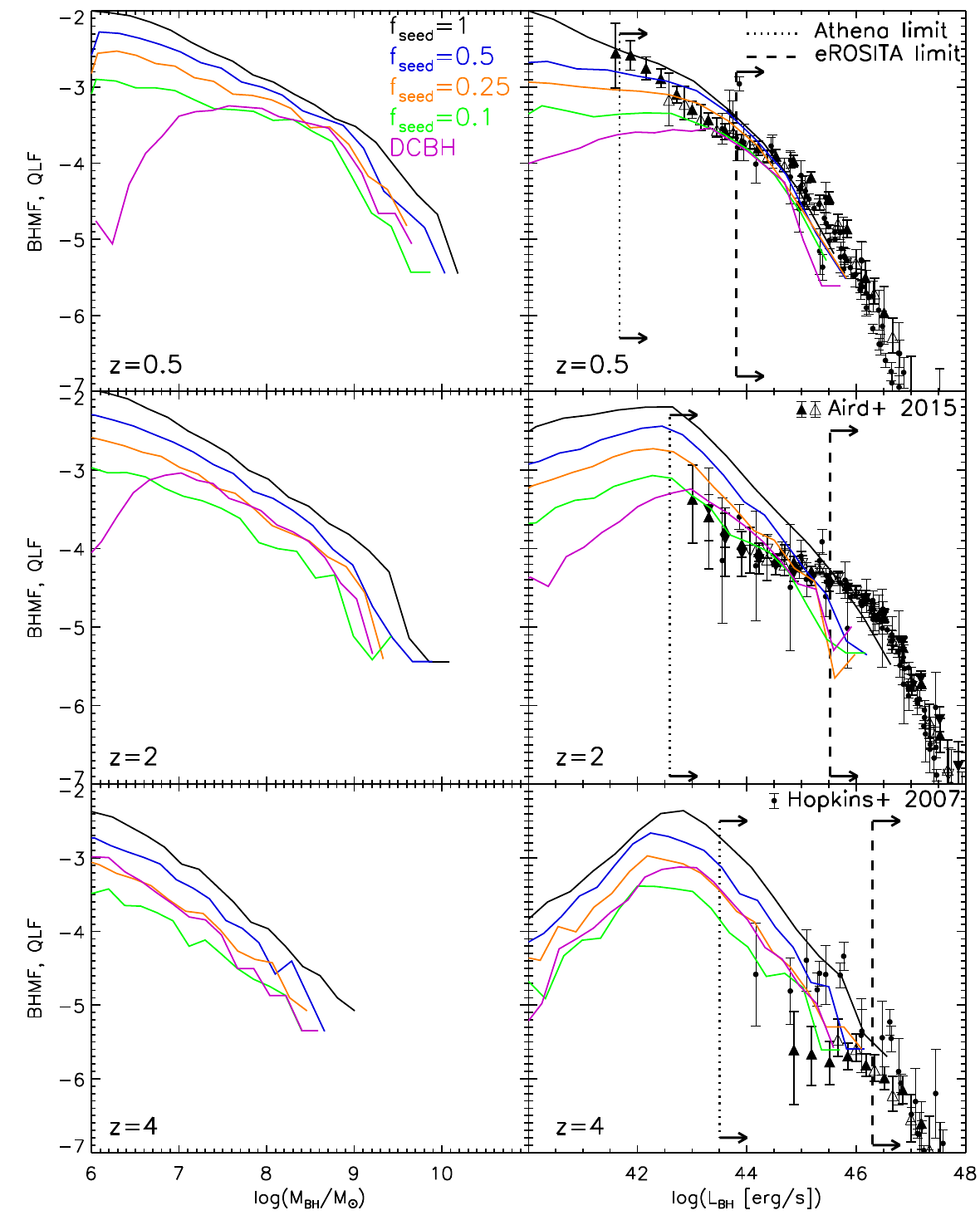}
    \caption{Black hole mass function (left) and bolometric luminosity
      function (right) for different $f_{\rm{seed}}$ models  and the DCBH seeding model
      at $z = 0.5, 2$ and $4$, with observational QLF data points (black points -
      \citeauthor{Hopkins2007} \citeyear{Hopkins2007}; open triangles - hard
      X-ray data from \citeauthor{Aird2015} \citeyear{Aird2015}; closed
      triangles - soft X-ray data from \citeauthor{Aird2015}
      \citeyear{Aird2015}; all converted to bolometric luminosities). Note that at $z = 2$ we show two bins of X-ray
      data from \citet{Aird2015} (upward triangle for $z = 1.5-2$ and downward
      triangle for $z = 2-2.5$, noting they are qualitatively very similar).
      We also show the sensitivity limits for both Athena (dotted line) and
      eROSITA (dashed line). Stricter seed conditions result in lower mass and
      luminosity functions, especially at the low-end.} 
    \label{fig:QLF_BHMF}
\end{figure*}

Furthermore, we note that the masses and baryon fractions of galaxies hosting
Illustris seeds tend to be relatively redshift independent (though $M_*$ does
increase slightly with time), since Illustris always seeds haloes as they
cross a fixed halo mass threshold ($M_{\rm{halo}} > 5 \times 10^{10} \:
\mathrm{ h^{-1} M_\odot}$). In contrast, in our DCBH-based seeding we find
that at low redshift (below $z \sim 3$), only galaxies with relatively low
baryon fractions are seeded, as they are more likely to have the
low-metallicity necessary to prevent excessive fragmentation of gas.  

The top left-hand panel of Figure~\ref{fig:seednumbers} shows the total number of galaxies and haloes above
several mass thresholds as a function of redshift (galaxies - solid lines; haloes - dot-dashed lines), showing that
the number of high-mass galaxies grows monotonically with time, and the number
of low-mass galaxies peaks at moderate redshift (below which we find a slight
decrease due to increasing number of galaxy mergers).  In particular, we note
that the dot-dashed blue line ($> 5 \times 10^{10} \: \mathrm{h^{-1} M_\odot}$) is
the number of haloes above the mass threshold for black hole seeding in the
original Illustris simulation, and thus corresponds roughly to the number of
black holes seeded in Illustris (neglecting mergers).  We compare
these total halo numbers to the number of galaxies satisfying our criteria
for DCBH seed formation at redshift $z$ (dotted line), and the cumulative
number of galaxies satisfying the same seedability criterion at any earlier
time (dashed line).  This shows that the total number of seeded black holes by
$z = 0.5$ is comparable between the original Illustris simulation (blue line)
and our DCBH-based model (dashed black line); however, where the original
simulation has a monotonically increasing seeded number, our DCBH seed model
peaks at $z \sim 8$ where the conditions for direct collapse are most common.
Furthermore, we note that our total number of DCBH seeds is actually higher than the original simulation.  This increase
is due to the lower host galaxy mass (see also Figure \ref{fig:host_mass_vs_z}), and thus represents a population of black holes in lower-mass galaxies than included in the Illustris black hole population.  Given how little the scaling relation is affected, this would represent a correspondingly low-$M_{\rm{BH}}$ population and would thus have minimal impact on the population of resolved black holes we consider in this analysis (see also discussion in Section \ref{sec:globalproperties}).

From this figure, we find a very large number of galaxies which could
potentially form DCBH seeds.  Based on our model criteria (see
Section~\ref{sec:DCBH_seeding}), we predict $\sim 10^{-2} \: \:
\mathrm{cMpc^{-3}}$ possible DCBH seed candidates by $z \sim 6$, far above the
number density of high-redshift quasars \citep[e.g.][]{Fan2004,
  Jiang2016}. However, we note that the majority of these objects will not
grow significantly.  In fact, at $z \sim 5$, we find $\sim 9 \times 10^{-4}
\mathrm{cMpc^{-3}}$ possible DCBH seeds whose host halo will have reached the
mass threshold for seeding in Illustris, and our regrowth calculations suggest
only $\sim 10^{-5} \mathrm{cMpc^{-3}}$ will have reached $M_{\rm{BH}} = 10^7
\mathrm{M_\odot}$. This is still well above the observed quasar number
density, so incorporating additional seed criteria such as a Lyman-Werner
radiation threshold or a gas inflow cut \citep[e.g.][]{Wise2019} would be a
worthwhile exercise, bearing in mind a possibility that only a fraction of
DCBH seeds grows sufficiently to power high redshift quasars.

In the lower left-hand panel of Figure~\ref{fig:seednumbers} we plot the
fraction of 
galaxies at redshift $z$ (with $M_{\rm{gal}} > 3 \times 10^9 \:
\mathrm{M_\odot}$) which would have formed a DCBH seed at some earlier time,
according to our DCBH model.  We find that the seeded fraction reaches a peak
of $\sim 25\%$ at $z \sim 8$, while at lower redshifts the fraction drops,
suggesting only $\sim 10\%$ of local galaxies above $3 \times 10^9 \mathrm{M_\odot}$ should have a black hole formed
through direct collapse. We also note a slight upturn in the seeded fraction
below $z \sim 3$.  However, this increase is not due to more direct collapse
seeds; rather it is a result of the decrease in $N_{\rm{gal}} (M > 3 \times
10^9 \: \mathrm{M_\odot})$ caused by low redshift galaxy mergers (seen in the
red curve of the upper panel). 

\begin{figure}
    \centering
    \includegraphics[width=8.5cm]{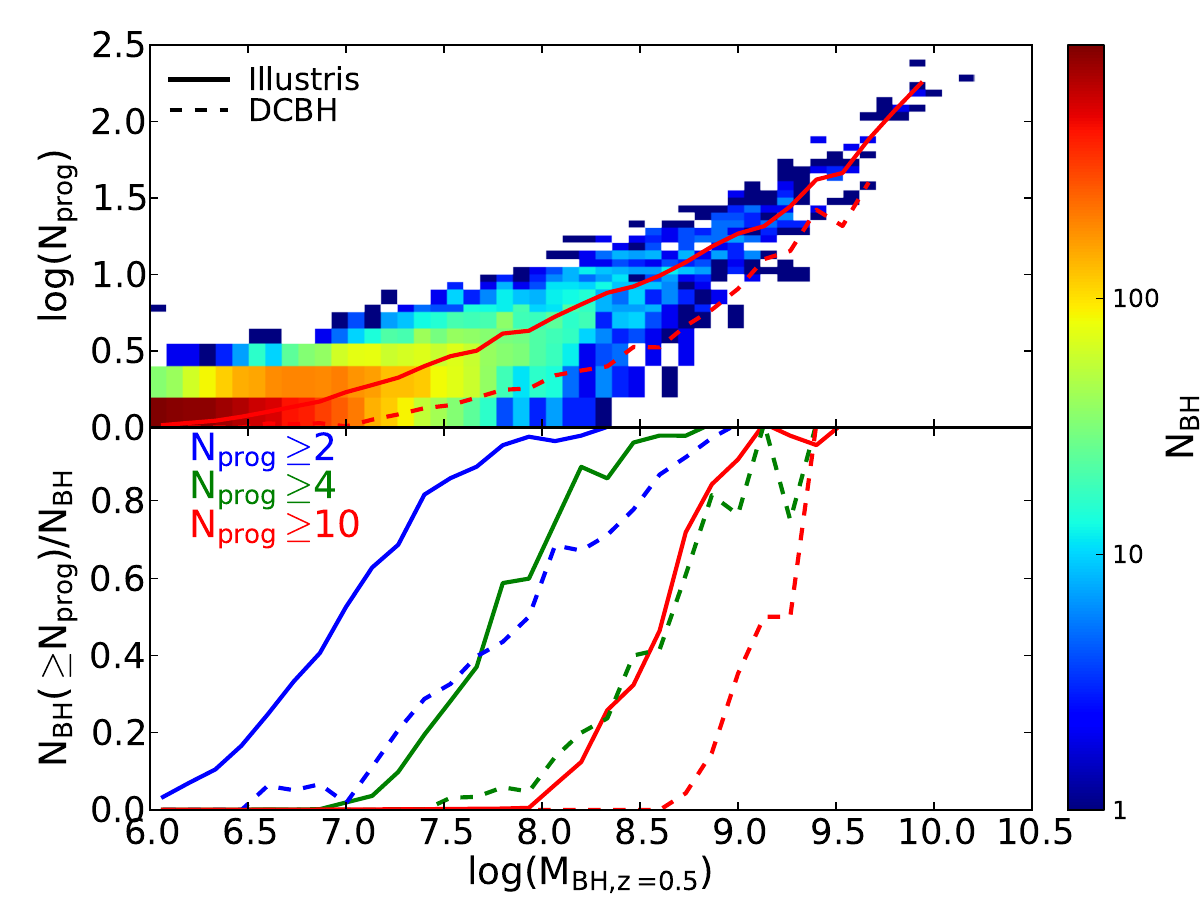}
    \caption{\textit{Top:} Distribution of the number of progenitor black
      holes for each Illustris black hole at $z = 0.5$, as a function of its mass. The red lines show $\left <N_{\rm{prog}} \right >$ as a function of $M_{\rm{BH}}$ for the original simulation (solid) and the DCBH model (dashed).  \textit{Bottom:} Fraction of Illustris black holes with at least 2 (blue), 4, (green), and 10 (red) progenitor black holes (corresponding to $f_{\rm{seed}}=0.5, 0.25, 0.1$, respectively).}
    \label{fig:nprogs}
\end{figure}

In the right-hand panel of Figure~\ref{fig:seednumbers} we plot the fraction
of original Illustris black holes which are seeded according to several
alternate seeding models (i.e. $N_{\rm{BH,new}}/N_{\rm{BH,Illustris}}$).  In
particular, we consider seeding based on gas spin only (see
Section~\ref{sec:spin_seeding}), having selected a maximum gas spin value
corresponding to approximately 
50, 25, and 10 per cent of Illustris black hole seeds
($\lambda_{\rm{g},50}$, $\lambda_{\rm{g},25}$, and $\lambda_{\rm{g},10}$, respectively),
and for our DCBH seed model.  We also
show the total number of black holes seeded in Illustris for reference.  Consistent with Figure~\ref{fig:spin_vs_z} which showed $\lambda$ had
minimal z-dependence, seeding according to instantaneous galaxy spin alone is
roughly redshift-independent, similar to the simpler
$f_{\rm{seed}}$-based seeding (not plotted, but would correspond to horizontal
lines by construction).  In contrast to this, DCBH seeding grows with time
until $z \sim 4$.  Below this the fraction of seeded black holes decreases as
few new DCBH seeds form and existing black holes proceed to merge together.

This peak at $z \sim 4$ is qualitatively different from the left-hand panel of
Figure~\ref{fig:seednumbers}, which demonstrated that the peak in DCBH seeding
occurs at $z \sim 8$.  This difference is in the definition of what is being
plotted: left-hand panel shows the fraction of galaxies which satisfy the
criteria for seeding a black hole at redshift $z$, while the right-hand panel shows the fraction of Illustris black holes which have at least one progenitor which satisfied our seeding criteria at some earlier time.  At lower-redshift, a given Illustris black hole has a larger number of progenitor galaxies than at higher redshift, and is thus more likely to have at least one progenitor satisfying the necessary criteria, hence the increase in seedable fraction.  At lower redshifts ($z \lesssim 4$), however, most of the progenitor galaxies
capable of forming a DCBH have already merged into a large enough halo for
Illustris to insert a black hole; thus most black holes inserted into Illustris at at $z \lesssim 4$ tend not to have any DCBH-seedable progenitor galaxies making low-$z$ DCBH formation very rare.

In Figure~\ref{fig:finalmass_vs_z} we show the distribution of black hole mass at $z = 0.5$ as a function of redshift at which the black hole is seeded ($z_{\rm{seed}}$) for the original Illustris simulation (top) and our DCBH seed model (middle).  We define $z_{\rm{seed}}$ as the seed redshift of the most massive progenitor within the given black hole merger tree, rather than the earliest seed redshift within the tree, noting that the most massive progenitor is often not the first to be seeded (e.g. see Figure~\ref{fig:progenitor_histories}), but is the most relevant quantity to consider here.

We find a direct correlation between $M_{\rm{BH}}(z=0.5)$ and $z_{\rm{seed}}$,
consistent between the original Illustris simulation, our DCBH seeding, and our $f_{\rm{seed}}$-based models (not plotted).  
As previously discussed, there are few low-$z_{\rm{seed}}$ black holes in the DCBH model.  This is further quantified in the bottom panel of Figure~\ref{fig:finalmass_vs_z} where we plot the fraction of original Illustris black holes which should be seeded according to our DCBH model.  This shows a rapid decline in the fraction of Illustris black holes with $z_{\rm{seed}} \lesssim 4$ which should form via our DCBH model.  

\subsection{Global black hole properties}
\label{sec:globalproperties}

Having considered the galaxies in which black holes should form, we use the
full recalculated growth history of all black holes to investigate the impact
the seeding model has on measurements of the full black hole population. In
particular, in Figure~\ref{fig:QLF_BHMF} we show the black hole mass function
(BHMF) and
quasar luminosity function (QLF) from the original Illustris simulation, for our
fractional seed models and for our DCBH-based
seeding model. We also show observational data from \citet[][black
  points]{Hopkins2007} and \citet[][orange triangles]{Aird2015}, and the
sensitivity limits of the Athena \citep{ATHENA2013} and eROSITA
\citep{eROSITA2012} surveys.  For the \citet{Aird2015} data, the Athena and
eROSITA limits, we convert from X-ray to bolometric luminosities using the SED
of \citep{Hopkins2007}.

The BHMF for the fractional seed models shows generally expected behavior: as
$f_{\rm{seed}}$ decreases, the BHMF decreases.  At high redshift, this shift
in BHMF is proportional to the decrease in $f_{\rm{seed}}$, as the lower
$f_{\rm{seed}}$ decreases the number of seeded black holes without affecting
the growth of those black holes which are seeded.  At low redshift, however,
black hole mergers become significant for high-mass black holes, and thus the
post-seed growth behaviour depends on seeding probability as well.  Since
massive, low-redshift black holes tend to have a large number of progenitor
black holes, most of the high-$M_{\rm{BH}}$ black holes are seeded regardless
of $f_{\rm{seed}}$. 

We show this explicitly in Figure~\ref{fig:nprogs}, the upper panel of which
shows the correlation between the mass of each black hole at $z = 0.5$,
$M_{\rm{BH}}(z = 0.5)$, and the total number of progenitor seeds it had,
$N_{\rm{prog}}$, while the lower panel shows the fraction of black holes which
have at least 2, 4, or 10 progenitors (corresponding to $f_{\rm{seed}}$ = 0.5,
0.25, 0.1, respectively). This shows that $\left <N_{\rm{prog}} \right >$
grows with $M_{\rm{BH}}$, and the mass scale above which we would expect most
black holes to have at least one progenitor seed would be $M_{\rm{BH}}(z =
0.5) \sim 1.5 \times 10^7 \: \mathrm{M_\odot}$, $6 \times 10^7 \:
\mathrm{M_\odot}$, and $4 \times 10^8 \: \mathrm{M_\odot}$ for $f_{\rm{seed}}
= 0.5, 0.25$, and $0.1$, respectively.  We further note that these scales are
redshift-dependent, as higher redshift black holes tend to have fewer
progenitors. In addition to the original Illustris data, we show the number of
progenitor black holes in our DCBH seeding model using dashed lines.  As
expected, $N_{\rm{prog}}$ is lower in this model, as fewer black holes are
seeded. 

The decrease in $M_{\rm{BH}}$ among high-mass black holes is due to the
decreased mass gained via mergers (since fewer progenitors were seeded).  This is further compounded by a lower accretion rate, both because $M_{\rm{BH}} \propto
M_{\rm{BH}}^2$ (Equation~\ref{eq:accretion}), and the higher accretion rate in the original simulation produced stronger feedback and thus lower gas density and higher temperature than we should have in the new seed model, which each contribute to a lower $\dot{\mathrm{M}}$. Furthermore, the most massive
black holes require a relatively early seed in a host with sufficient gas
density to fuel an extended phase of near-Eddington growth (e.g. the most
massive progenitor curves in Figure~\ref{fig:progenitor_histories}).  When the
seed fraction is decreased, there are fewer chances and thus a generally
delayed onset of Eddington growth, producing a lower black hole mass.  The net
effect is that the number of high-mass black holes tends to be roughly
unchanged but with slightly lower mass (as seen in
Figure~\ref{fig:progenitor_histories}), producing a slightly steeper slope in
the BHMF. We further note that this also means that although the high-end mass function is lower in our DCBH model, the black hole occupation fraction is still high in massive galaxies (rapidly approaching 1 for total galaxy masses $\sim \rm{few} \times 10^{12} M_\odot$). Conversely, low mass black holes with single progenitors are less likely to be seeded, but those that do have unchanged masses. Thus the decrease in the low-end BHMF does represent a comparable decrease in the black hole occupation fraction among low-mass galaxies as has also been found to be the case when seeding at lower masses based on local gas properties \citep[see][]{Habouzit2017}.  

The BHMF for the DCBH model is roughly consistent with the $f_{\rm{seed}} \sim
0.25$ model, except for the low-redshift, low-$M_{\rm{BH}}$ end, where the
DCBH seed model falls off drastically (e.g. below $M_{\rm{BH}} \sim 10^7 \mathrm
{M_\odot}$ at $z = 2$) due to the difference in redshifts at which black hole
seeds form.  In particular, most seeds in the DCBH model form at an early time
(see Figure~\ref{fig:seednumbers}), in contrast to $f_{\rm{seed}}$ based
models which continue seeding black holes at all redshifts as is the case in
the original Illustris model.  Since the DCBH
model seeds very few black holes at low redshift and $M_{\rm{BH}}(z = 0.5)$ is
directly correlated with $z_{\rm{seed}}$, nearly all black holes have had
sufficient time to reach at least $\sim 10^{6.5} \: \mathrm{M_\odot}$ (given
our accretion prescription), and
thus the mass function decreases rapidly below this scale. Thus our model
predicts very few low redshift, low mass black holes which formed via direct
collapse route. However, we note that this does not necessarily mean that there
will be a turnover in the overall observed BHMF. Our model only considers
DCBHs, but we also expect black holes may form via NSCs or from PopIII
remnants. Black holes formed through these mechanisms may contribute
significantly to the mass function, providing the potential to partially or
even completely overcome the turnover found in the DCBH
population. Alternatively, or additionally, it is possible that some of
the direct collapse seeds will not grow by accretion as efficiently as in our
model which may lead to a less significant decrease in the low mass-end BHMF
shape. Furthermore, our analysis does not include black holes which may have been seeded by direct-collapse but whose host halo never reaches the threshold for seeding in the original simulation ($5 \times 10^{10} \: h^{-1} \: \mathrm{M_\odot}$).  As discussed in Section \ref{sec:seedproperties}, this will represent a low-$M_{\rm{BH}}$ population, and thus should have minimal impact on the results shown here.  In particular, we note that the mean mass for black holes in haloes below $10^{11} \: \mathrm{M_\odot}$ in the original simulation is only $2.6 \times 10^5 \: \mathrm{M_\odot}$ at z = 0.5 and less than 3\% have $M_{\rm{BH}} > 10^6 \: \mathrm{M_\odot}$, suggesting these black holes will have minimal impact on the results considered here, where we generally limit our analysis to $M_{\rm{BH}} > 10^6 \: \mathrm{M_\odot}$.  This is supported by results of \citet{Wang2019}, who found that seeding black holes at earlier times with $M_{\rm{BH,seed}}=10^3 \: h^{-1} \: \mathrm{M_\odot}$ resulted in comparable mass black holes when reaching the Illustris-seed threshold (for haloes at $M_{\rm{DM}}=10^{10} \: h^{-1} \: \mathrm{M_\odot}$, $\langle log(M_{\rm{BH}}) \rangle \sim 5.18$ with standard deviation $\sim 0.55$).  \citet{Habouzit2017} also looked at low-$M_{\rm{BH,seed}}$ simulations, and found that the black holes would not grow to $10^6 \: \mathrm{M_\odot}$ until the host galaxy reached a stellar mass of a few $\times 10^8 \: \mathrm{M_\odot}$ (and a few $\times 10^9$ in simulations which used a delayed cooling model for SN feedback), broadly consistent with Illustris' halo mass threshold.

Similar to the BHMF, the QLF tends to exhibit a simple normalization shift
equivalent to $f_{\rm{seed}}$ at either high redshift or low-$L_{\rm{BH}}$,
the regimes in which most black holes have undergone few or no mergers and
thus modifying the seed prescription changes which black holes form but
does not affect their subsequent growth.  As with the high-mass BHMF, the low
redshift, high-luminosity black holes tend to have undergone numerous mergers,
and thus the number is generally unaffected by the seed model.  However, we
find that the low-redshift, high-end of the QLF is somewhat less affected by
the seed 
prescription than the BHMF is, such that current observational data cannot
differentiate between our seed formation models. This is broadly consistent with
the semi-analytic model of \citet{VolonteriLodatoNatarajan2008}, who also
found the mass function to be more sensitive to seed model than the luminosity
function.

\begin{figure}
    \centering
    \includegraphics[width=9cm]{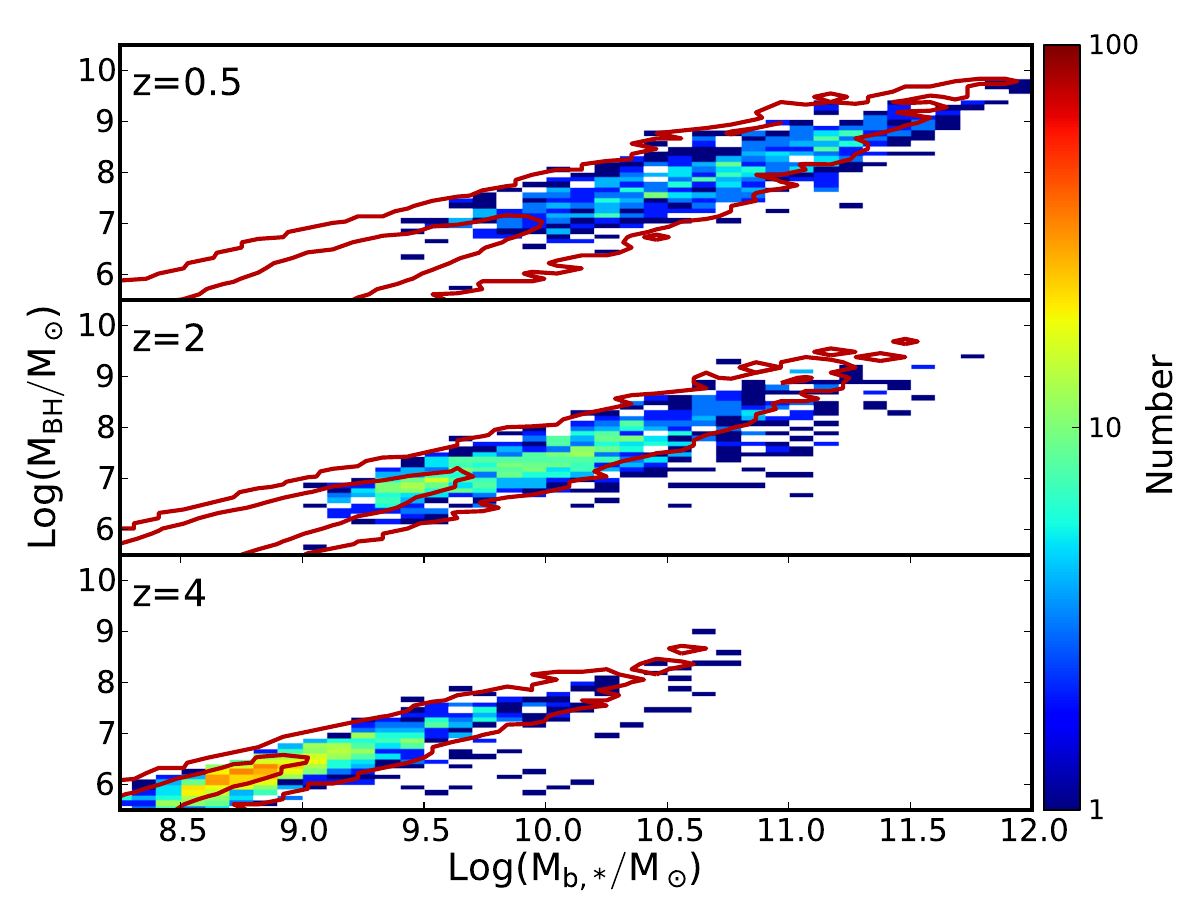}
    \caption{Scaling relation between black hole mass ($M_{\rm{BH}}$) and
      bulge stellar mass ($M_{\rm{b,*}}$, approximated using the stellar
      half-mass radius) at $z = 0.5$, $2$ and $4$.  Red contours show the relation
      from the original Illustris simulation, while the colourmap shows the
      distribution from the DCBH model. For $z < 3$, the low mass-end of the
      scaling relation becomes less and less populated in the DCBH model as
      seeding efficiency drops with redshift and low mass black holes have
      sufficient time to grow significantly in mass.}
    \label{fig:scaling_relation}
\end{figure}

We note that we find good agreement with observational data across a wide
range of luminosities and redshifts.  At low redshift and high luminosities,
all of our seed models reproduce the observed data well (though noting that
the simulations are
volume-limited, and thus do not produce the brightest quasars). At the
lowest luminosities the rarer seed models under-predict the luminosity
function, but additional seed models which provide lower-mass and
lower-redshift seeds would be expected to fill this gap in.  At moderate
redshifts ($z \sim 2$) we find that the rarest seed models (including the DCBH
model) match the faint end most effectively. Although the brightest objects
are underestimated, we expect this to be at least partially due to the limited
boxsize and somewhat overestimating the impact of feedback in our DCBH seed
model (see the discussion of the scaling relation in
Figure~\ref{fig:scaling_relation}, below).  At the highest redshifts ($z \sim
4$), all our models remain broadly consistent with observations, though this
is largely due to the uncertainty in the observed high-redshift faint-end
luminosity function. We note that the upcoming eROSITA data will not help
differentiate between the different black hole seeding models (see redshift
dependent eROSITA limits), which will only become possible (both at high and
low redshifts) with the Wide Field Imager of the Athena mission.

\begin{figure}
    \centering
    \includegraphics[width=8.5cm]{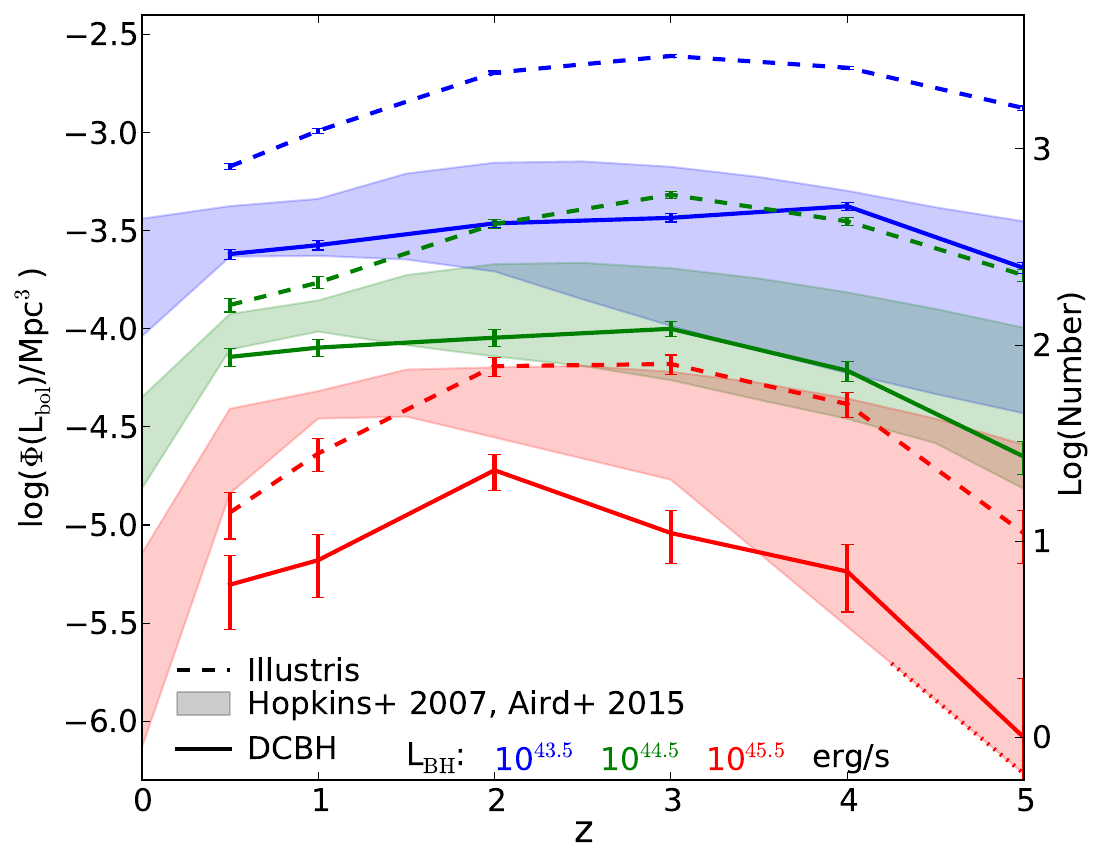}
    \caption{Comoving number density of AGN for three bins in bolometric
      luminosity (blue - $10^{43}-10^{44} {\rm erg\,s^{-1}}$; green -
      $10^{44}-10^{45} {\rm erg\,s^{-1}}$; red - $10^{45}-10^{46} {\rm erg\,s^{-1}}$) from the original Illustris simulation (dashed lines), our DCBH model (solid lines), and observational constraints (shaded regions represent region spanned by multiple fitting models from \citeauthor{Hopkins2007} \citeyear{Hopkins2007} and data from \citeauthor{Aird2015} \citeyear{Aird2015}, see text for details). We find that the DCBH model improves on the Illustris model except for the brightest AGN, where our model is least well-constrained.}
    \label{fig:number_density}
\end{figure}

In Figure~\ref{fig:number_density} we show the total comoving number density
of AGN binned by bolometric luminosity to investigate cosmic downsizing,
comparing the original Illustris simulation (dashed lines) to the DCBH model
(solid lines).  We also show observational constraints in the form of shaded
regions, showing the areas spanned by the data from \citep{Aird2015} and a
range of models fitted to observational data \citep[the `Full', `PLE',
  `Scatter', `Schechter', and `LDDE' fitting models from][]{Hopkins2007}.  We
find that Illustris overpredictss low-luminosity AGN
($10^{43}-10^{44} {\rm erg\,s^{-1}}$, blue) at high-redshift, whereas the DCBH
model more closely matches the observed data.  Similarly, at intermediate
luminosities ($10^{44}-10^{45} {\rm erg\,s^{-1}}$, green) the DCBH model
matches observations very closely, where the original simulation over-predicted
the number density, especially at high-redshift.  In the brightest sample
($10^{45}-10^{46} {\rm erg\,s^{-1}}$, red) the DCBH under-predicts the number
density at all but the highest redshifts.  However, we note that the brightest AGN are least
well-constrained in our model, as they are the objects where the AGN feedback
(which is unaffected by our DCBH model) plays the strongest role, which we can better understand by looking at the relation between black holes and their host galaxies.
Furthermore, we emphasize that the comoving number densities from \citet{Hopkins2007} are based on analytic fits to the observed data.  The observed data includes significant scatter (see Figure~\ref{fig:QLF_BHMF}) and can vary significantly depending on the fitting model used, especially when extrapolating toward faint luminosities.

In Figure~\ref{fig:scaling_relation} we show the impact the DCBH seeding
method has on the scaling relation between the stellar mass of a galaxy
($M_{\rm{b,*}}$; defined as the stellar mass within the stellar half-mass
radius of the subhalo) and the corresponding $M_{\rm{BH}}$ (the mass of the
central black hole within the galaxy) for the original Illustris simulation
(contours) and our new DCBH model (colourmaps). At high redshift we see
excellent agreement between the two, with the exception of some
low-$M_{\rm{BH}}$ outliers in the DCBH model.  These outliers are caused when
the black hole which undergoes the fastest growth in the original simulation
is not seeded. Instead there is a lower-mass black hole that grows up toward
the main relation, but will not reach it until a lower redshift.

At lower redshift, we note that the DCBH model only fills the high-mass end of
the scaling relation, without any low-mass objects \citep[similar to][who also
  found that stricter seed models result in a significant fraction of low-mass
  galaxies which do not host black holes]{VolonteriLodatoNatarajan2008}.  This
lack of a low-mass end is due to the low-redshift, low-mass galaxies not satisfying
the conditions for direct-collapse seed formation while those that are seeded
did so at sufficiently high-redshift, providing enough time for both the black
hole and the galaxy to grow to at least a moderate mass scale\footnote{We
  stress again this result may also be sensitive to the accretion model.  If a
  less-efficient accretion model were adopted, the slower growth of black
  holes seeded at high-redshift would also lead to a larger population of
  low-mass black holes at low redshift.}. We emphasize that this suggests that
low-redshift, low-mass galaxies rarely host massive black holes formed
\textit{via direct collapse}; however they may still host SMBHs formed through
other channels, e.g. from NSCs or PopIII stellar remnants.  We also note that
the threshold for seeding in these low-mass galaxies may be sensitive to the
metal enrichment model in Illustris \citep[see Section~\ref{sec:DCBH_seeding}
  and][]{Vogelsberger2013}. Thus the precise mass scale below which we do not
find DCBHs may be model-dependent, but the qualitative result should hold.

\begin{figure}
    \centering
    \includegraphics[width=9cm]{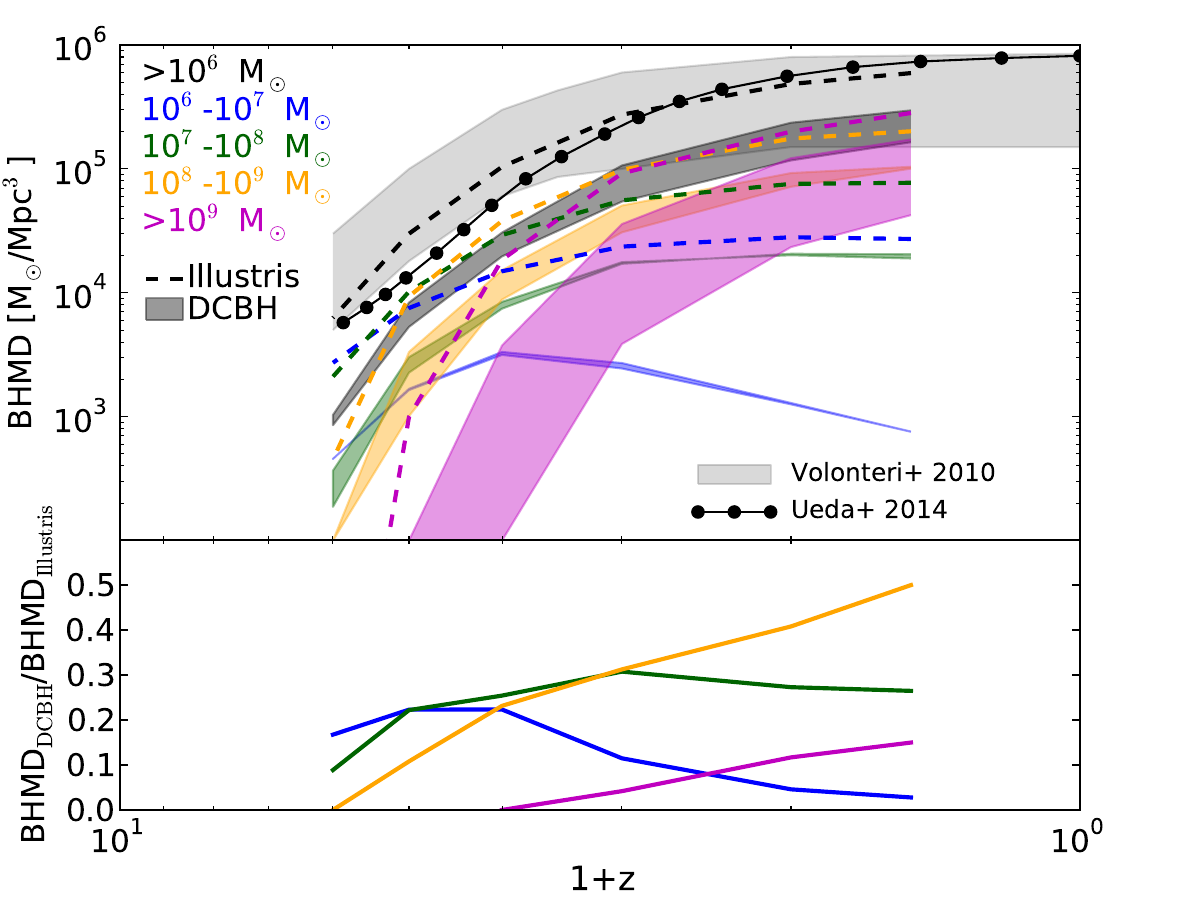}
    \caption{\textit{Top:} black hole mass density from the original Illustris
      simulation (dashed line) and our DCBH model (shaded regions), binned by
      $M_{\rm{BH}}$.  Each shaded region spans the range of densities based on
      our post-processed $M_{\rm{BH}}$ to the densities if each $M_{\rm{BH}}$
      is rescaled based on the scaling relation shift (see main text for more
      details). \textit{Bottom:} ratio between DCBH mass density and the
      original Illustris mass density for different mass bins.}
    \label{fig:BHMD}
\end{figure}

In addition, we note that below $z \sim 2$ the high-mass end in the DCBH model
lies below the relation from the original simulation.  As discussed earlier in
this section, black holes in the DCBH model often have lower masses than in
the original simulation (see e.g. the mass function in
Figure~\ref{fig:QLF_BHMF}). However, we note that one caveat of our model is
that the feedback from the original simulation remains unchanged, and thus the
growth of massive black holes in the DCBH model may be underestimated, as the
gas supply around black holes which reach the self-regulated regime in the
original simulation has been impacted by a stronger feedback than it would be
with the lower-mass DCBH.  As such, while we do expect the DCBH model to have
a lower high-$M_*$ end, our results provide a lower bound rather than a direct
prediction.

\begin{figure*}
    \centering
    \includegraphics[width=16cm]{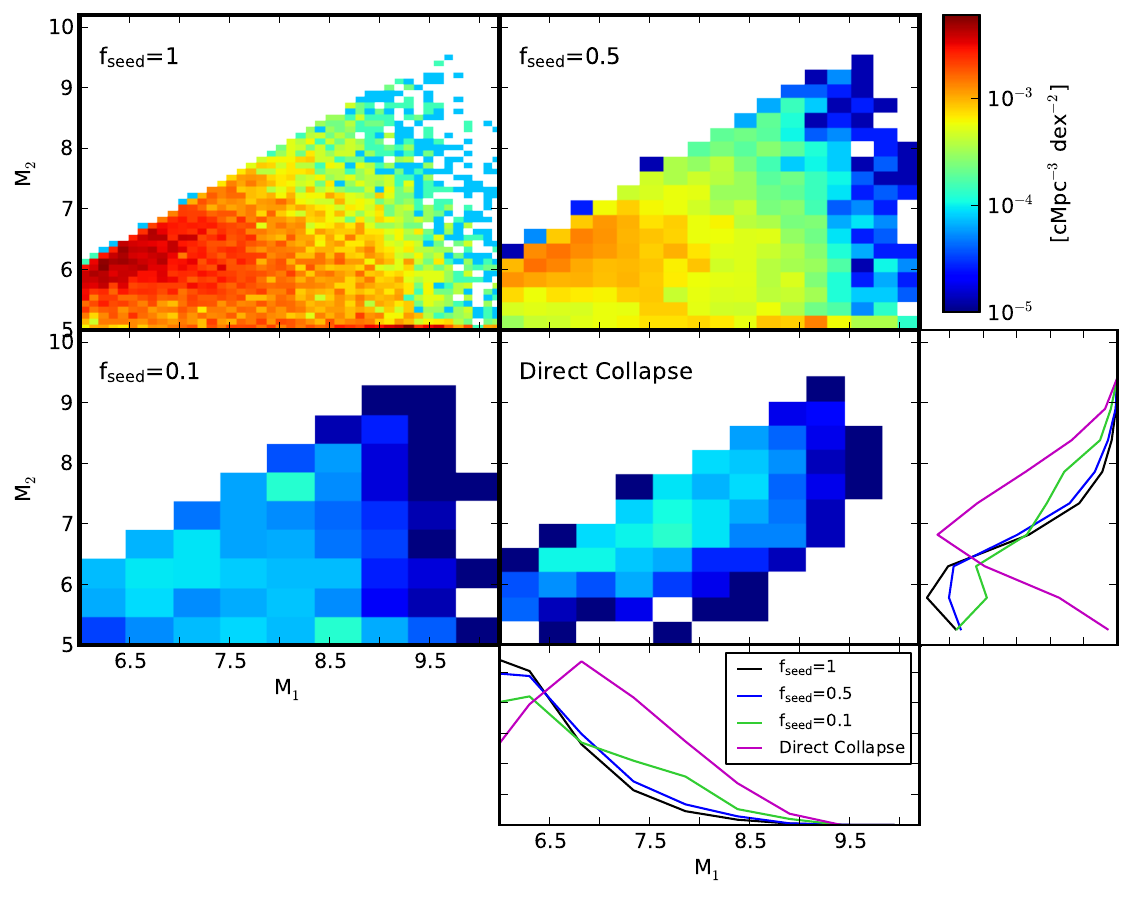}
    \caption{Number density of mergers as a function of $M_1$ (more massive
      black hole) and $M_2$ (less massive black hole), for $f_{\rm{seed}}=1, 0.5, 0.1$, and the DCBH seed model. Changing $f_{\rm{seed}}$ primarily impacts the overall merger rate with a small effect on the distribution.  The DCBH model, however, produces a substantially different distribution, lacking the highest mass-ratio mergers.}
    \label{fig:mergerrate_2dhist}
\end{figure*}

We also consider the global black hole mass density as a function of redshift
for a range of $M_{\rm{BH}}$-bins in the upper panel of Figure~\ref{fig:BHMD}.
As noted above, high-$M_{\rm{BH}}$ are likely somewhat under-massive due to
overestimation of the effect of feedback. In this figure we hence compare the
original Illustris mass densities (dashed lines) to a range of possible values
in our DCBH model (shaded region). The shaded areas show the region bounded
by two constraints: the explicit calculation of $M_{\rm{BH}}$ from our DCBH
model, and an adjusted value to account for this overestimated feedback.  For the
adjusted mass, we compare the best-fitting relation for the scaling relation
(see Figure~\ref{fig:scaling_relation}) of the original simulation and the
DCBH model, and at the high-mass end where the DCBH model is lower, we
increase $M_{\rm{BH}}$ of the DCBH by the difference between the two
relations.  The lower-$M_{\rm{BH}}$ bound is thus expected to be a lower limit
as the black hole growth has been underestimated. Similarly, the
higher-$M_{\rm{BH}}$ bound is likely the upper limit, as we expect the DCBHs to have a somewhat lower mass than the original simulation due to the less
frequent seeding.  As such, the shaded regions represent a well-defined
constraint for the mass densities. The width of the shaded regions is clearly
mass dependent, with only the most massive black holes above $10^9 \, {\rm
  M_\odot}$ being affected significantly by the overestimated feedback in our
DCBH model.

In the lower panel of Figure~\ref{fig:BHMD} we show the ratio of the DCBH mass
density to that of the original Illustris simulation. Here we see that the
density of low-mass black holes ($<10^7 \: \rm{M}_\odot$) tends to drop off below
z$\sim$3-4, as new seeds rarely form and thus black holes tend to grow out of
the lowest mass bin. For higher mass black holes, we see that at high-redshift
we have much lower mass densities in the DCBH model, but as time passes the
more efficient growth starts to increase the mass density up toward that of
the original Illustris simulation.

\begin{figure*}
    \centering
    \includegraphics[width=16.0cm]{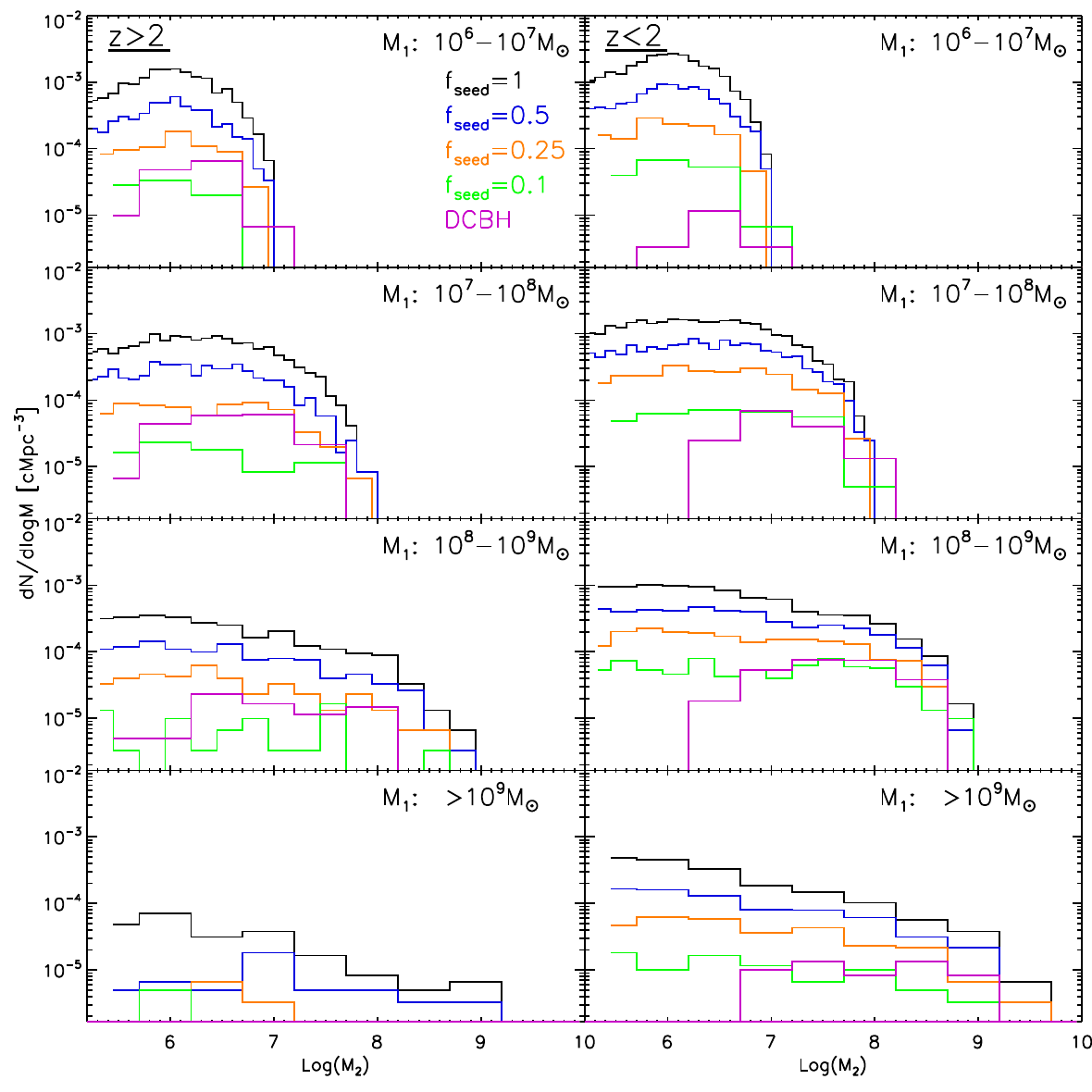}
    \caption{Number density of mergers binned by primary black hole mass for different seed models (colour), for mergers at $z > 2$ (left) and $z < 2$ (right). The DCBH model results in fewer mergers overall, particularly those involving low-mass black holes at low-redshift.}
    \label{fig:mergerrate_1dhist}
\end{figure*}

From Figure~\ref{fig:BHMD} several interesting features can be deduced. First,
consistent with our findings above, black hole mass density of the DCBH model
is systematically lower with respect to the Illustris prediction. 
Second, by splitting the contribution of total black hole mass density into different mass
bins, we see that at high redshift the density is dominated by low-mass black holes.
Hence, the discrepancy with regard to current high-$z$ observational estimates for black hole
mass density is likely due to underestimating this low-mass population, requiring either a slightly higher DCBH seed efficiency (but see our discussion above that our DCBH seeding model is likely rather conservatively high), or other channels of black hole formation contributing to the total mass density (e.g. SMBHs from PopIII or NSC seeds).  Furthermore, at low redshift the overall density is dominated by high mass black holes, suggesting that, if the growth of SMBHs from low mass seeds is negligible, matching current observational densities would require more efficient growth than our post-processing analysis finds, possibly up to including some super-Eddington phases.  It will be interesting to see if future JWST, Athena, IPTA, and LISA observations will sufficiently inform us about these growth channels to break these degeneracies in theoretical models \citep[e.g., see][]{Sesana2007, Plowman2010, Sesana2011, Pacucci2015, Klein2016, Natarajan2017, Valiante2018}.

\subsection{Black hole merger rates}
\label{sec:mergers}

In addition to the impact on the overall black hole population and its cosmic
growth, changing the seeding prescription strongly modifies the rate of black hole mergers, as well
as the mass ratio of the merging black holes.  

Naively one might expect that the number of mergers $N_{\rm{merger}}$ scales roughly with $f_{\rm{seed}}^2$, 
if $f_{\rm{seed}}$ were the probability of any given black hole having formed.
However, this ignores the full history of the merger trees: more massive black
holes tend to have undergone a large number of mergers (see
Figure~\ref{fig:nprogs} and the discussion in
Section~\ref{sec:globalproperties}), and thus have a much higher probability
of at least one of those progenitors having been seeded.  For this reason,
mergers involving more massive black holes will be less sensitive to the
seeding prescription with the number of mergers in the DCBH model becoming
significantly closer to the Illustris prediction for higher mass black holes.

\begin{figure*}
    \centering
    \includegraphics[width=17cm]{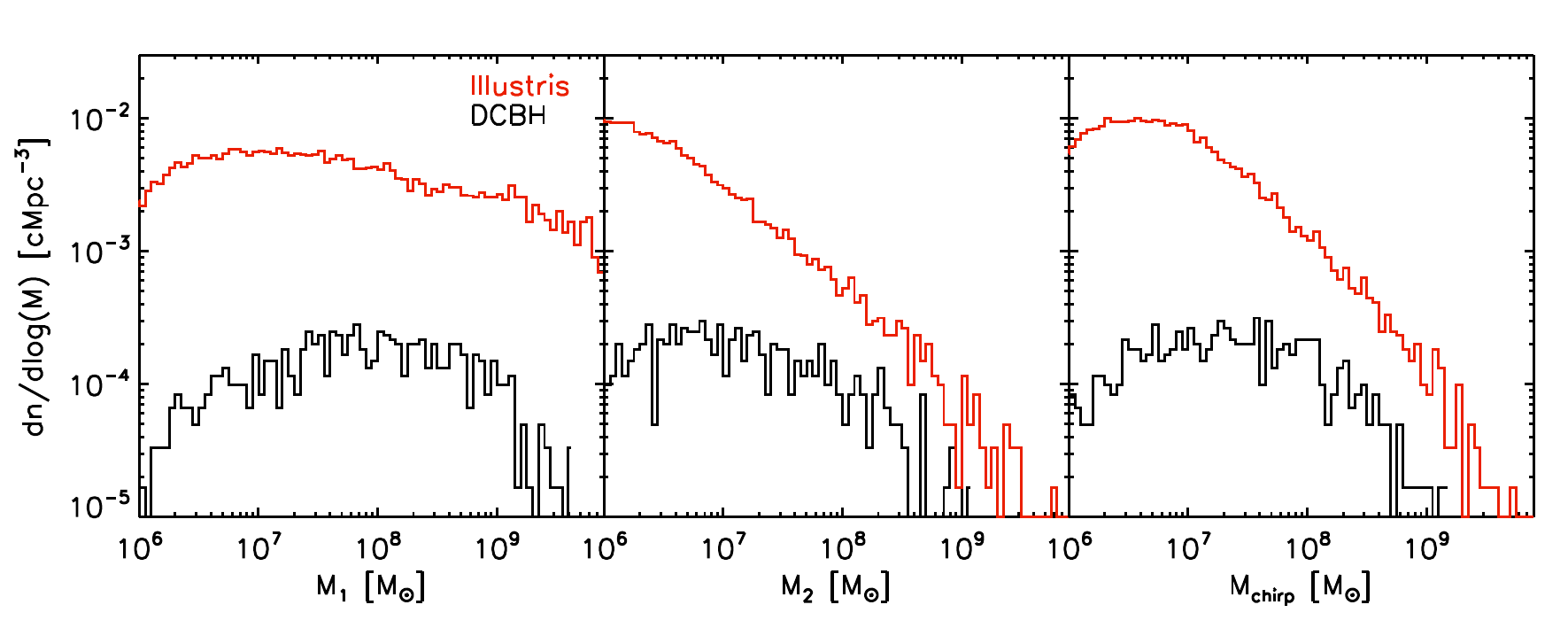}
    \caption{Number density of mergers as a function of $M_1$ (left), $M_2$
      (middle), and chirp mass ($M_{\rm{chirp}}$, right), for both original
      Illustris (red) and our new DCBH model (black).  In addition to the
      overall merger number, the strongest impact is on the secondary black
      hole masses for any given merger, where the DCBH model has a much higher
      typical value than the original Illustris simulation.}
    \label{fig:nmergers}
\end{figure*}

In Figure~\ref{fig:mergerrate_2dhist} we plot the 2-dimensional distribution
of the number density of mergers, binned by both $M_1$ and $M_2$ (the masses
of the more massive and less massive black holes involved in the merger,
respectively).  For fixed $f_{\rm{seed}}$ models, we find that the merger
frequency peaks at $M_1 \sim 10^{6.3} \: \rm{M}_{\rm \odot}$ and $M_2 \sim 10^{6}
\: \rm{M}_{\rm \odot}$.  The overall $M_1 - M_2$ distribution appears roughly comparable (barring an overall normalization shift), but we do note that the lower-mass mergers are affected more strongly by a lower $f_{\rm{seed}}$, so lower $f_{\rm{seed}}$ results in a distribution weighted slightly more toward high-mass mergers. 

The DCBH based seeding model, however, produces a qualitatively different
merger-mass distribution which peaks at $M_1 \sim 10^{7.75} \: \rm{M}_{\odot}$,
$M_2 \sim 10^{7} \: \rm{M}_{\rm \odot}$ and generally has $M_2$ within $\sim 1$ dex
of $M_1$.  We also see this quite clearly in the 1D distributions of $M_1$ and
$M_2$ in Figure~\ref{fig:mergerrate_2dhist}, which show that mergers in the
DCBH model peak at both higher $M_1$ and $M_2$ (see also Figure~\ref{fig:mergerrate_1dhist} and~\ref{fig:nmergers}).

We also note that an earlier analysis of the {\small EAGLE} simulation found merger rates dominated by low-mass black holes, but much more strongly peaked toward seed mass black holes.  They also found that decreasing the black hole seed mass to $M_{\rm{BH,seed}} \sim 10^4 \: \mathrm{M_\odot}$ (but maintaining the halo threshold for seed formation) dramatically affected the merger distribution, eliminating nearly all mergers except those with $M_2 \sim M_{\rm{BH,seed}}$ and a small population of $M_1, M_2 \sim 10^8 \: \mathrm{M_\odot}$, reinforcing the importance of improved seed models when looking at merger rates \citep{Salcido2016}.

In Figure~\ref{fig:mergerrate_1dhist} we plot 1-dimensional histograms of
$M_2$ for various bins of $M_1$. There is some slight $M_2$-dependence, with
high-$M_2$ values tending to be less sensitive to $f_{\rm{seed}}$ due to the
increase in progenitor black holes. Other than this small effect, the
distribution is qualitatively similar for all $f_{\rm{seed}}$ models beyond an
overall normalization shift.  However, as in
Figure~\ref{fig:mergerrate_2dhist}, the DCBH-based seeding model shows a
dramatically different distribution, dropping off rapidly at low $M_2$.  This
dropoff is consistent with the dropoff in mass function seen in
Figure~\ref{fig:QLF_BHMF}, which shows dramatically fewer low-mass black holes
in the DCBH model at low redshifts.  Because the DCBH seeding model has most
black holes form at relatively high redshift, at lower redshift there are far
fewer low-$M_{\rm{BH}}$ black holes available to undergo mergers, and thus we
find fewer low-$M_2$ mergers.  This also means that the low-$M_2$ dropoff is
partially a redshift-dependent factor, and so high redshift mergers exhibit a
less extreme dropoff (see the left- vs. right- panels of
Figure~\ref{fig:mergerrate_1dhist}).

\begin{figure*}
    \centering
    \includegraphics[width=15cm]{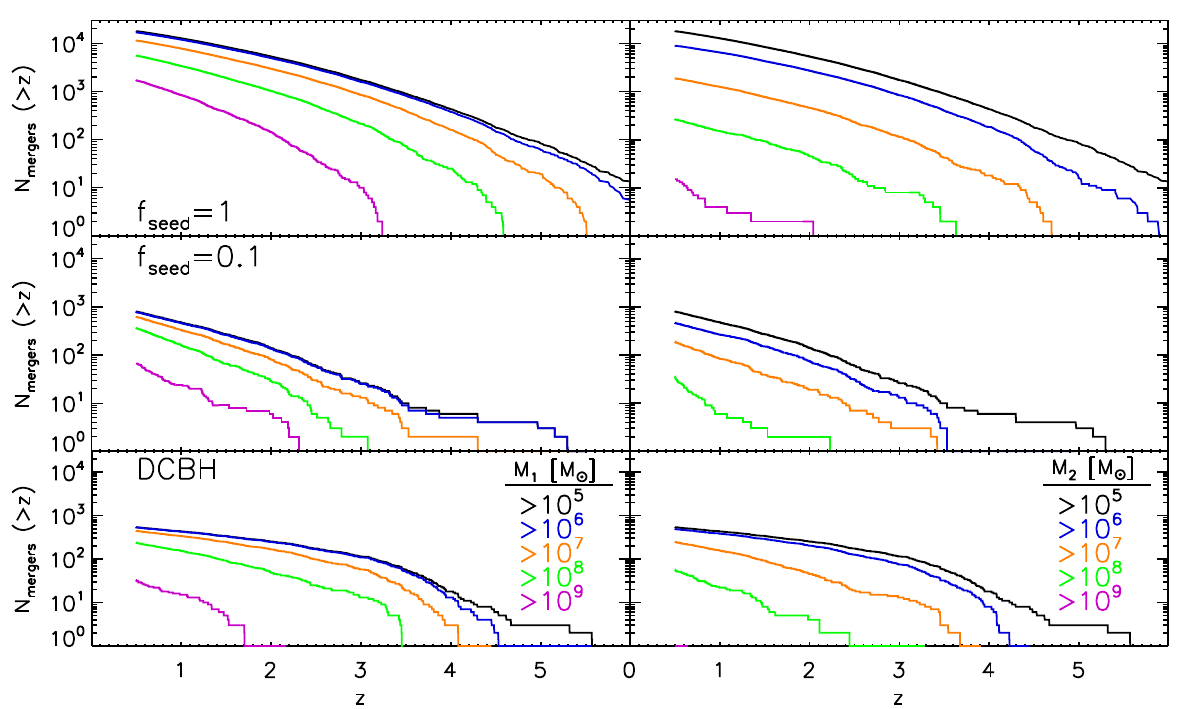}
    \caption{Cumulative merger number binned by primary (left) and secondary
      (right) black hole mass for $f_{\rm{seed}}=1$ (top), $f_{\rm{seed}}=0.1$
      (middle), and the DCBH seed model (bottom).  Changing $f_{\rm{seed}}$
      impacts the normalization but the curves are otherwise qualitatively
      similar.  The DCBH model, however, produces very few mergers at low redshift.}
    \label{fig:mergerplot_primarysecondary}
\end{figure*}

In Figure~\ref{fig:nmergers} we show the number density of mergers for
original Illustris and our DCBH model as a function of primary mass ($M_1$,
the more massive black hole involved in the merger; left panel) and secondary
mass ($M_2$, the less massive black hole; center panel).  In addition to a significant decrease in merger number, we again see that the DCBH model tends to have much higher secondary black hole masses, while the primary masses tend to be relatively similar (though without reaching as high- or low- masses).  We note that the Illustris simulation predicts that $M_1$ should peak near $\sim 10^7 \: \mathrm{M_\odot}$, broadly consistent with the findings from the {\small EAGLE} simulation, though {\small EAGLE} produced more mergers with $M_2 \sim M_{\rm{BH,seed}}$ and relatively fewer intermediate-$M_2$ mergers \citep{Salcido2016}. In addition, we show (right panel) the merger number as a function of chirp mass, a measure of the effective mass of a binary system affecting the gravitational wave signal of the merger, defined as
\begin{equation}
M_{\rm{chirp}}=\frac{(M_1 M_2)^{3/5}}{(M_1 + M_2)^{1/5}} \:.
\label{eq:mchirp}
\end{equation}
We see that the DCBH model peaks at a higher $M_{\rm{chirp}}$ ($\sim
10^7-10^8 \: \rm{M}_{\rm \odot}$, compared to $\sim 10^6-10^7 \: \rm{M}_{\rm \odot}$ from the original
Illustris simulation), and does not extend to as low values (ending just below
$10^6 \: \rm{M}_{\rm \odot}$), suggesting DCBH-DCBH mergers are most common at
slightly higher masses then the original Illustris simulation predicts. It is
worth emphasizing that the merger
density of our DCBH model remains consistent with predicted constraints from
Pulsar Timing Array \citep[e.g.][]{Middleton2016}, though the current PTA
constraints remain very broad, hence future tighter constraints would help
differentiating the models.

In Figure~\ref{fig:mergerplot_primarysecondary} we look at the redshift
distribution of mergers, showing the cumulative number of mergers before
redshift $z$, binned by $M_1$ (left panels) and $M_2$ (right panels), for
$f_{\rm{seed}}=1$ (top), $f_{\rm{seed}}=0.1$ (middle), and the DCBH model
(bottom).  Again we find the $f_{\rm{seed}}$ models are qualitatively similar
to one another, but the DCBH seed model is significantly different.  Mergers
with low-$M_1$ in the DCBH seed model tend to occur exclusively at high
redshifts. Most DCBH mergers with $M_1 < 10^7 \: \rm{M}_{\rm \odot}$ occur by $z
\sim 3.5$, whereas in $f_{\rm{seed}}$-based models $N_{\rm{mergers}}$ grows
with time throughout the simulation. Similarly, there are few $M_2 < 10^6 \:
\rm{M}_{\rm \odot}$ DCBH mergers below $z \sim 3.5$, though $10^6 \: \rm{M}_{\rm \odot}
< M_2 < 10^7 \: \rm{M}_{\rm \odot}$ do continue to lower redshifts.  

\begin{figure}
    \centering
    \includegraphics[width=8.5cm]{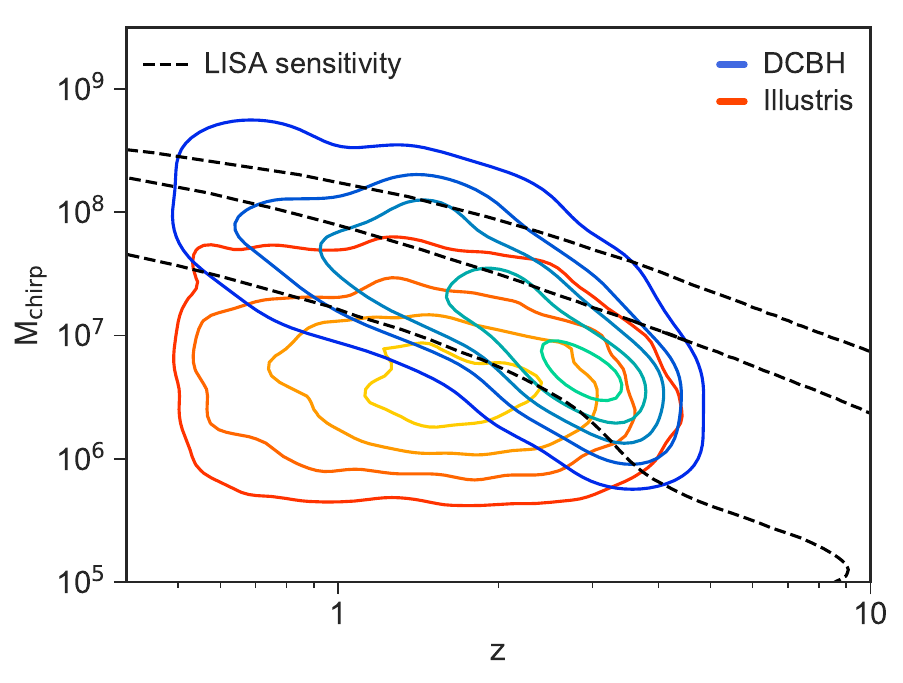}
    \caption{Evolution of merger chirp mass as a function of redshift, for the
      original Illustris simulation (orange contours) and our DCBH model (blue
      contours).  Continuous seeding makes the Illustris simulation have
      roughly redshift-independent chirp masses, while the lack of low redshift seeding
      in the DCBH model results in a substantial increase in $M_{\rm{chirp}}$
      with time. Black dashed lines show LISA sensitivity for fixed Signal-to-Noise ratios of 10, 50, and 500 (top, middle, and bottom, respectively), showing that the redshift evolution of chirp mass should be detectable by the LISA mission.}
    \label{fig:mchirp_vs_z}
\end{figure}

In Figure~\ref{fig:mchirp_vs_z} we show the relation between chirp mass and
merger redshift.  In the original Illustris simulation (orange contours), we see that
the typical $M_{\rm{chirp}}$ remains roughly constant with time. In these models, black holes are seeded all the way to $z = 0$, so at all redshifts we find mergers spanning the full range of masses (see also Figure~\ref{fig:mergerplot_primarysecondary}, which shows comparable increases in $N_{\rm{mergers}}$ for all $M_{\rm{BH}}$). In contrast, however, the DCBH mergers (blue contours) show a strong increase in $M_{\rm{chirp}}$ with decreasing redshift, as expected from our earlier analysis: at low-redshift, low mass black holes tend to be much rarer (Figure~\ref{fig:QLF_BHMF}), resulting in far fewer low-mass mergers (the flattening of the $N_{\rm{mergers}}$ for low-$M_{\rm{BH}}$ in Figure~\ref{fig:mergerplot_primarysecondary}).  Thus the redshift distribution of $M_{\rm{chirp}}$ detections can provide a strong means of discriminating between seed models.

\begin{figure}
    \centering
    \includegraphics[width=8.5cm]{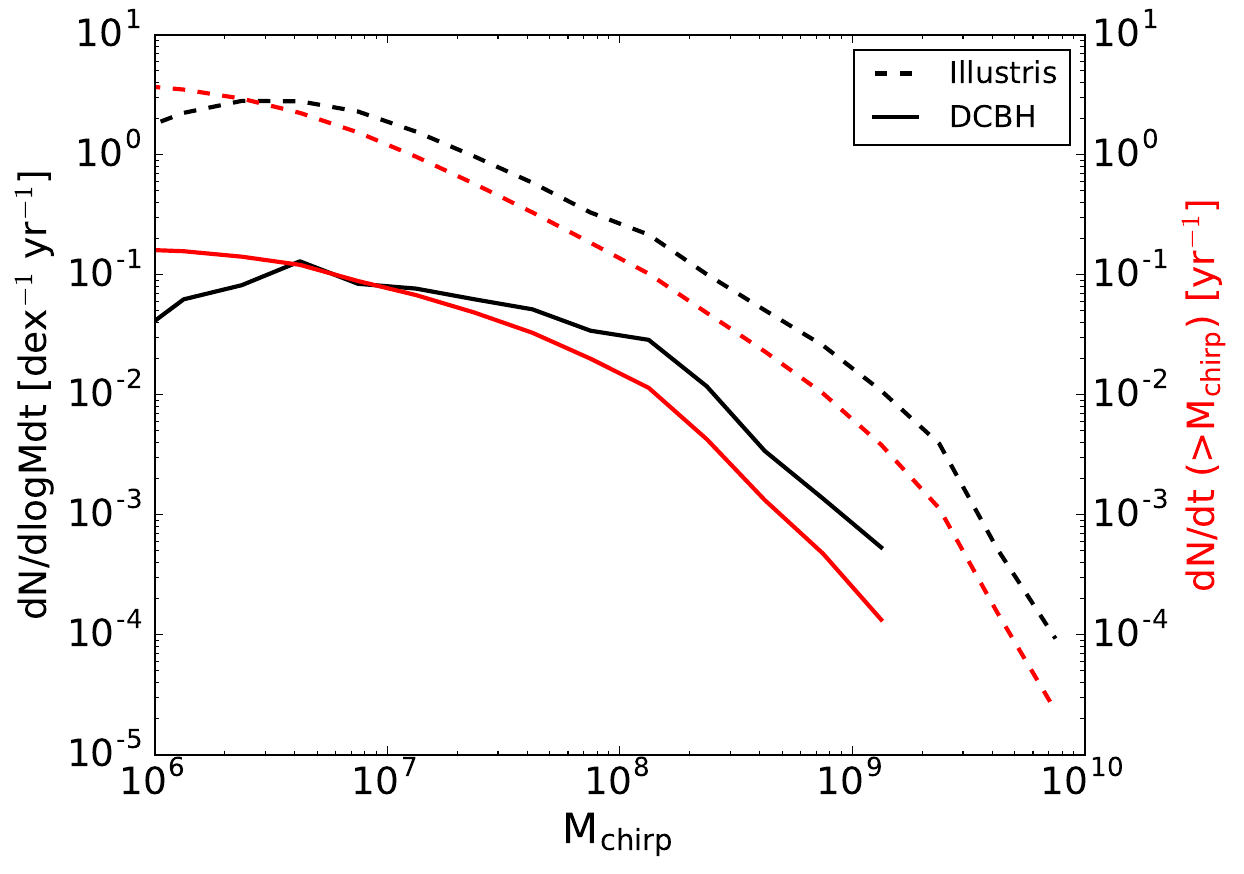}
    \caption{The rate of black hole mergers over cosmic time predicted by Illustris (dashed lines) and the DCBH model (solid lines) as a function of chirp mass, for bins of $M_{\rm{chirp}}$ (black) and cumulatively for $>M_{\rm{chirp}}$ (red), showing the substantial decrease in merger rates in the DCBH model. }
    \label{fig:mergerrate}
\end{figure}

Finally, in Figure~\ref{fig:mergerrate} we consider the rate at which mergers
occur throughout the observable universe, both at a given chirp mass
($dN/dlogMdt$, black curves) and the cumulative number above a given chirp mass
($dN/dt$, red curves).  Here we see that the merger rate peaks for a chirp
mass of $\sim 3-4 \times 10^6 \: \rm{M}_{\rm \odot}$ (original Illustris) and $\sim
4 \times 10^6 \: \rm{M}_{\rm \odot}$ (DCBH model).  For the Illustris simulation,
this peak is comparable to the peak seen in Figure~\ref{fig:nmergers}, as we
would expect; although the observable volume is redshift-dependent,
$M_{\rm{chirp}}$ is approximately independent of redshift (see
Figure~\ref{fig:mchirp_vs_z}), so the merger rate is comparable to the merger
density.  However, for the DCBH model, $M_{\rm{chirp}}$ evolves strongly with
redshift, shifting the merger rate toward a lower chirp mass. From the
cumulative rate of mergers above a given chirp mass (red curves), we see that
the DCBH model predicts $\sim 1 \: dex$ fewer mergers, such that we would only
expect to detect $\sim 0.1$ supermassive black hole merger per year.  However, we note that this rate is only for
mergers involving black holes which are well resolved by the simulation, so
detectors sensitive to lower-mass ranges can be expected to make more
detections than our simulation predicts. 

For example, we note that LISA \citep{AmaroSeoane2017} has primary sensitivity
which extends down to $\sim 10^4 \: \rm{M}_{\rm \odot}$, well below our limit of $\sim
10^6 \: \rm{M}_{\rm \odot}$.  Given the steepness of the halo mass function (and the
associated increase of low mass galaxies, see Figure~\ref{fig:seednumbers}),
we note that the actual number of detectable mergers will likely be higher as
a result of the lower-mass mergers unresolved by our simulation 
This is consistent with previous analyses \citep[e.g. ][]{Sesana2004, Sesana2007, TanakaHaiman2009, Sesana2011, Ricarte2018}, which have found that the distribution of merger rates peaks at low black hole masses, tending toward the black hole seed mass used in the model.
However, at higher
masses where we resolve well the black holes, we expect our estimates to be much closer to complete.

We also note that this predicted rate assumes that black holes form \textit{only} via direct collapse,
and the incorporation of black holes formed via alternative seed pathways (e.g. PopIII remnants or in NSCs)
would be expected to significantly increase the overall detection rate.  In
particular, a recent paper by \citet{Dayal2019} used a semi-analytic model to
predict high redshift merger rates, and found that for $z > 5$ mergers, LISA
detections of mergers between a DCBH and a stellar mass seed would outnumber
the DCBH-DCBH merger detections by two orders of magnitude or
more. Furthermore, similar to \citet{Ricarte2018}, \citet{Dayal2019} found
that the merger rate would peak at very low masses, with LISA-detectable
mergers (based on SNR $> 7$) peaking at $M_{\rm{BH}}(1+z) \sim 10^4-10^5
\: \rm{M}_{\rm \odot}$, and \citet{Bellovary2019} found that black hole mergers in dwarf galaxies (below the masses considered in this work) will also be detectable by LISA suggesting that the predictions we make for the rate of high-mass DCBH-DCBH mergers represents only a small subset of all LISA detections. 

As such, our predicted merger rate can be treated as a lower-limit for
detections. However, this is dependent on the treatment of merging timescales.
In particular, the black hole model in Illustris repositions black holes
toward the potential minimum of its host halo to prevent N-body scattering
with dark matter or stellar particles, which we note could result in
underestimating the migration time for a secondary black hole brought into a
halo by a galaxy merger.  Furthermore, the simulation merges black holes as
soon as they are within a smoothing length of one another \citep[see
  Section~\ref{sec:growth} and ][]{Sijacki2015}, although the actual hardening
timescale may be quite significant \citep[reaching up to $\sim 10^9$ yr for
  major mergers with $M \sim 10^9 \: \rm{M}_{\rm \odot}$, depending on orbital
  parameters; see, e.g.][]{Kelley2017}. 

Adding long hardening timescales would substantially delay the merger of any given black hole pair, decreasing the rate of high-redshift mergers which may be detected.  However, it would be expected to produce a corresponding increase in late-time merger rates, and the lower-redshift at which the merger occurs would make them more easily detectable. Thus we expect that a more realistic hardening timescale would not substantially impact the detection rate of black hole mergers, only that it might lead to a distribution weighted more toward low-redshift (and we further note that an initial investigation suggests that incorporating a delay of $\sim$1 Gyr would not significantly impact the overall merger rates, consistent with findings of \citeauthor{Salcido2016} \citeyear{Salcido2016}).  

\section{Conclusions}
\label{sec:conclusions}

In this work, we have developed a novel method to investigate the impact black
hole seeding models have on the cosmic evolution of the entire black hole
population. Once a given seed model is assumed, our method recalculates entire
black hole growth and merging histories taking advantage of the Illustris cosmological simulation of galaxy
formation. Hence our method presents a more self-consistent and comprehensive
approach than the traditional semi-analytic models while at the same time
allowing us to test a large number of seeding models which would be
computationally prohibitively expensive with full hydrodynamical simulations
of structure formation.  Recall the caveat to this approach is that the impact of AGN feedback on the gas supply around the black hole is unchanged during the post-processing approach; thus for black holes with lower mass compared to the original simulation our model may over-estimate the effect of AGN feedback and thus decrease the predicted accretion rate.  However, this effect will generally be significant only for a small fraction of black holes, at the highest mass end where feedback is strongest (for details, see Section \ref{sec:globalproperties} and Figure \ref{fig:BHMD}).  

In our analysis we test stochastic and physically-motivated seed models based on the direct collapse black hole (DCBH) formation scenario, find the impact on low- and high-redshift black hole populations and merger rates, and determine which observables provide a means for distinguishing between simulated models for seed formation.  From this, we draw the following conclusions:

\begin{itemize}
\item Our novel tool produces post-processed black hole growth histories based on the simulation data available at each timestep. These histories match the original simulation when using the same seeding parameters, but alternate conditions for seeding can also be easily incorporated at a negligible fraction of the original simulation's computational expense.  In particular, we implement simple stochastic seeding models (our $f_{\rm{seed}}$-based models) and a physically motivated seeding model which uses galaxy merger trees to find progenitor galaxies in atomically cooling haloes with low metallicity and low central gas spin in which may be the forming sites of black hole seeds via direct collapse (our DCBH model).

\item Galaxy properties of newly-seeded black holes in the original Illustris simulation are largely redshift independent (though with some increase in metallicity below $z \sim 2$), and seeding continues throughout cosmic time.  In contrast, our DCBH seed model is biased toward higher redshifts.  Furthermore, the low redshift seeds (below $z \sim 3.5$) tend to be found in galaxies with relatively low baryon fractions, which have a tendency for low-metallicity while still having enough gas to form DCBH seeds.

\item About $10-20\%$ of all simulated galaxies with total mass above $3 \times 10^9 \: \rm{M}_\odot$ have a progenitor which satisfies all our criteria for direct collapse formation, suggesting only $\sim 20\%$ of Illustris black holes may have formed in this manner.

\item All of our seed models (down to $f_{\rm{seed}}=0.1$) produce quasar
  luminosity functions (QLFs) consistent with current observational
  constraints.  Except at very high redshift (where the observed QLF is poorly
  constrained), the bright end of the QLF ($L_{\rm{BH}} > 10^{43.5}$ erg s$^{-1}$) is
  similar for all seed models.  The faint-end QLF shows significant dependence
  on seeding, however, particularly at low redshift, suggesting that upcoming
  surveys such as Athena should be able to put constraints on the seeding
  model. We note, however, that this is based on a single-formation mechanism
  via direct collapse. Additional formation pathways, such as PopIII remnants
  and NSCs, can provide additional black hole populations, particularly at
  lower masses.

\item The scaling relation between black hole and host galaxy masses is
  strongly affected by the seeding model assumed, with the DCBH seed model
  resulting in very few low black hole mass objects at low redshift, in contrast to the original Illustris simulation.  Hence, future observational campaigns that aim to probe the low mass end of the scaling relations may provide much needed insight into the likely nature of black hole seeding.

\item Total black hole mass density is sensitive to the seeding model adopted,
  and our results 
  imply that either DCBH seed need to undergo more efficient growth to match current observational constraints (possibly including a phase of super-Eddington
  accretion) or that the other seed
  formation channels lead to non-neglible cosmic black hole mass
  growth.  Future JWST, Athena, LISA, and PTA observations may be able to shed light on this degeneracy \citep[see also][for how LISA could differentiate between light and heavy seed models]{Sesana2007, Plowman2010, Sesana2011, Klein2016}.

\item Black hole - black hole mergers are dramatically affected by the seeding
  model.  The total number of mergers scales more strongly than linearly with
  $f_{\rm{seed}}$, with the DCBH model expecting approximately one order of
  magnitude fewer mergers than the original Illustris simulation.
  Furthermore, the expected merger frequency is mass-dependent, with mergers
  involving the lowest masses depending most strongly on the seeding
  frequency. Thus high mass-ratio mergers are most reduced within our DCBH seed model: black holes tend to reach the highest masses only at low-redshift, and the lack of low redshift seeding means there are very few low-mass DCBHs at late times with which the large black hole can merge.

\item The redshift evolution of merger chirp mass is strongly dependent on the
  seed formation model as well.  The continued seeding across all redshifts in
  the original simulation results in a roughly redshift-independent
  distribution of chirp masses.  In contrast, the lack of low-redshift seed
  formation in our DCBH model results in a typical chirp mass that strongly
  increases with time, reaching $\sim 10^8\, \rm{M}_{\rm \odot}$ at low-$z$, compared to $\sim
  10^{6.5}\, M_{\rm \odot}$ in the original Illustris simulation.

\end{itemize}

In summary, this work demonstrates that seed formation mechanisms can have significant impact on black hole populations across all mass and redshift ranges.
We have shown that differing seed models, including a more physically-motivated mechanism for seeding black holes in cosmological simulations based on the direct collapse scenario, are capable of broadly matching current observations across all redshifts. However, different seeding models have 
distinct predictions for higher redshift and lower mass black holes, which upcoming surveys such as JWST and Athena may be able to constrain. Perhaps most significantly, we have shown that seeding models have a strong effect on black hole merger rates, particularly at low redshift and high mass ratios, such that upcoming gravitational wave detections [e.g. LISA \citep{AmaroSeoane2017}, International Pulsar Timing Array \citep{Verbiest2016}, European Pulsar Timing Array \citep{Desvignes2016}, Parkes Pulsar Timing Array \citep{Reardon2016}, and NANOGrav \citep{Arzoumanian2018}] may be able to differentiate between seed formation pathways for supermassive black holes.

\section*{Acknowledgments}

We would like to thank Martin Haehnelt, Alberto Sesana, Marta Volonteri and
Priyamvada Natarajan for useful discussions and comments on this work and the referee for a helpful report. CD and DS acknowledge ERC
starting grant 638707 and support from the STFC. This research 
used: The Cambridge Service for Data Driven Discovery
(CSD3), part of which is operated by the University of Cambridge Research Computing on behalf of the STFC DiRAC
HPC Facility (www.dirac.ac.uk). The DiRAC component of
CSD3 was funded by BEIS capital funding via STFC capital grants ST/P002307/1 and ST/R002452/1 and STFC operations grant ST/R00689X/1. DiRAC is part of the National e-
Infrastructure. Simulations were
run on the Harvard Odyssey and CfA/ITC clusters, the Ranger and Stampede supercomputers at the Texas Advanced Computing Center as part of XSEDE, the Kraken
supercomputer at Oak Ridge National Laboratory as part
of XSEDE, the CURIE supercomputer at CEA/France as
part of PRACE project RA0844, and the SuperMUC computer at the Leibniz Computing Centre, Germany, as part
of project pr85je.

 \bibliographystyle{mn2e}       
 \bibliography{astrobibl}       

\begin{thebibliography}{110}
\expandafter\ifx\csname natexlab\endcsname\relax\def\natexlab#1{#1}\fi

\bibitem[{Agarwal} et~al.(2016){Agarwal}, {Smith}, {Glover}, {Natarajan} \&
  {Khochfar}]{Agarwal2016}
{Agarwal} B., {Smith} B., {Glover} S., {Natarajan} P., {Khochfar} S., 2016,
  \mnras, 459, 4209

\bibitem[{Aird} et~al.(2015){Aird}, {Coil}, {Georgakakis}, {Nandra}, {Barro} \&
  {P{\'e}rez-Gonz{\'a}lez}]{Aird2015}
{Aird} J., {Coil} A.~L., {Georgakakis} A., {Nandra} K., {Barro} G.,
  {P{\'e}rez-Gonz{\'a}lez} P.~G., 2015, \mnras, 451, 1892

\bibitem[{Amaro-Seoane} et~al.(2017){Amaro-Seoane}, {Audley}, {Babak}
  et~al.]{AmaroSeoane2017}
{Amaro-Seoane} P., et~al., 2017, arXiv: 1702.00786

\bibitem[{Arzoumanian} et~al.(2018){Arzoumanian}, {Brazier}, {Burke-Spolaor}
  et~al.]{Arzoumanian2018}
{Arzoumanian} Z., et~al., 2018, \apjs, 235, 2, 37

\bibitem[{Ba{\~n}ados} et~al.(2018){Ba{\~n}ados}, {Venemans}, {Mazzucchelli}
  et~al.]{Banados2018}
{Ba{\~n}ados} E., et~al., 2018, \nat, 553, 473

\bibitem[{Beckmann} et~al.(2017){Beckmann}, {Devriendt}, {Slyz}
  et~al.]{Beckmann2017}
{Beckmann} R.~S., et~al., 2017, \mnras, 472, 949

\bibitem[{Begelman} \& {Rees}(1978)]{BegelmanRees1978}
{Begelman} M.~C., {Rees} M.~J., 1978, \mnras, 185, 847

\bibitem[{Begelman} et~al.(2006){Begelman}, {Volonteri} \&
  {Rees}]{Begelman2006}
{Begelman} M.~C., {Volonteri} M., {Rees} M.~J., 2006, \mnras, 370, 289

\bibitem[{Bellovary} et~al.(2019){Bellovary}, {Cleary}, {Munshi}
  et~al.]{Bellovary2019}
{Bellovary} J.~M., {Cleary} C.~E., {Munshi} F., {Tremmel} M., {Christensen}
  C.~R., {Brooks} A., {Quinn} T.~R., 2019, \mnras, 482, 3, 2913

\bibitem[{Bondi}(1952)]{Bondi1952}
{Bondi} H., 1952, \mnras, 112, 195

\bibitem[{Bondi} \& {Hoyle}(1944)]{BondiHoyle1944}
{Bondi} H., {Hoyle} F., 1944, \mnras, 104, 273

\bibitem[{Bromm} \& {Loeb}(2003)]{BrommLoeb2003}
{Bromm} V., {Loeb} A., 2003, \apj, 596, 34

\bibitem[{Costa} et~al.(2014){Costa}, {Sijacki}, {Trenti} \&
  {Haehnelt}]{Costa2014}
{Costa} T., {Sijacki} D., {Trenti} M., {Haehnelt} M.~G., 2014, \mnras, 439,
  2146

\bibitem[{Curtis} \& {Sijacki}(2016)]{Curtis2016}
{Curtis} M., {Sijacki} D., 2016, \mnras, 457, L34

\bibitem[{Davis} et~al.(1985){Davis}, {Efstathiou}, {Frenk} \&
  {White}]{Davis1985}
{Davis} M., {Efstathiou} G., {Frenk} C.~S., {White} S.~D.~M., 1985, \apj, 292,
  371

\bibitem[{Dayal} et~al.(2019){Dayal}, {Rossi}, {Shiralilou}, {Piana},
  {Choudhury} \& {Volonteri}]{Dayal2019}
{Dayal} P., {Rossi} E.~M., {Shiralilou} B., {Piana} O., {Choudhury} T.~R.,
  {Volonteri} M., 2019, \mnras, 486, 2, 2336

\bibitem[{DeGraf} et~al.(2012){DeGraf}, {Di Matteo}, {Khandai} \&
  {Croft}]{DeGrafBHGrowth2012}
{DeGraf} C., {Di Matteo} T., {Khandai} N., {Croft} R., 2012, \apjl, 755, L8

\bibitem[{DeGraf} et~al.(2015){DeGraf}, {Di Matteo}, {Treu}, {Feng}, {Woo} \&
  {Park}]{DeGraf2015}
{DeGraf} C., {Di Matteo} T., {Treu} T., {Feng} Y., {Woo} J.-H., {Park} D.,
  2015, \mnras, 454, 913

\bibitem[{DeGraf} \& {Sijacki}(2017)]{DeGraf2017}
{DeGraf} C., {Sijacki} D., 2017, \mnras, 466, 3331

\bibitem[{Desvignes} et~al.(2016){Desvignes}, {Caballero}, {Lentati}
  et~al.]{Desvignes2016}
{Desvignes} G., et~al., 2016, \mnras, 458, 3, 3341

\bibitem[{Devecchi} \& {Volonteri}(2009)]{DevecchiVolonteri2009}
{Devecchi} B., {Volonteri} M., 2009, \apj, 694, 302

\bibitem[{Di Matteo} et~al.(2017){Di Matteo}, {Croft}, {Feng}, {Waters} \&
  {Wilkins}]{DiMatteo2017}
{Di Matteo} T., {Croft} R.~A.~C., {Feng} Y., {Waters} D., {Wilkins} S., 2017,
  \mnras, 467, 4243

\bibitem[{Di Matteo} et~al.(2012){Di Matteo}, {Khandai}, {DeGraf}
  et~al.]{DiMatteo2012}
{Di Matteo} T., {Khandai} N., {DeGraf} C., {Feng} Y., {Croft} R.~A.~C., {Lopez}
  J., {Springel} V., 2012, \apjl, 745, L29

\bibitem[{Di Matteo} et~al.(2005){Di Matteo}, {Springel} \&
  {Hernquist}]{DiMatteo2005}
{Di Matteo} T., {Springel} V., {Hernquist} L., 2005, Nature, 433, 604

\bibitem[{Dubois} et~al.(2014){Dubois}, {Pichon}, {Welker} et~al.]{Dubois2014}
{Dubois} Y., et~al., 2014, \mnras, 444, 1453

\bibitem[{Fan} et~al.(2004){Fan}, {Hennawi}, {Richards} et~al.]{Fan2004}
{Fan} X., et~al., 2004, AJ, 128, 515

\bibitem[{Fan} et~al.(2006){Fan}, {Strauss}, {Becker} et~al.]{Fan2006}
{Fan} X., et~al., 2006, AJ, 132, 117

\bibitem[{Feng} et~al.(2016){Feng}, {Di-Matteo}, {Croft}, {Bird}, {Battaglia}
  \& {Wilkins}]{Feng2016}
{Feng} Y., {Di-Matteo} T., {Croft} R.~A., {Bird} S., {Battaglia} N., {Wilkins}
  S., 2016, \mnras, 455, 2778

\bibitem[{Ferrarese}(2002)]{Ferrarese2002}
{Ferrarese} L., 2002, \apj, 578, 90

\bibitem[{Freitag} et~al.(2006{\natexlab{a}}){Freitag}, {G{\"u}rkan} \&
  {Rasio}]{Freitag2006b}
{Freitag} M., {G{\"u}rkan} M.~A., {Rasio} F.~A., 2006{\natexlab{a}}, \mnras,
  368, 141

\bibitem[{Freitag} et~al.(2006{\natexlab{b}}){Freitag}, {Rasio} \&
  {Baumgardt}]{Freitag2006a}
{Freitag} M., {Rasio} F.~A., {Baumgardt} H., 2006{\natexlab{b}}, \mnras, 368,
  121

\bibitem[{Fryer} et~al.(2001){Fryer}, {Woosley} \& {Heger}]{Fryer2001}
{Fryer} C.~L., {Woosley} S.~E., {Heger} A., 2001, \apj, 550, 372

\bibitem[{Gebhardt} et~al.(2000){Gebhardt}, {Bender}, {Bower}
  et~al.]{Gebhardt2000}
{Gebhardt} K., et~al., 2000, \apjl, 539, L13

\bibitem[{Genel} et~al.(2014){Genel}, {Vogelsberger}, {Springel}
  et~al.]{Genel2014}
{Genel} S., et~al., 2014, \mnras, 445, 175

\bibitem[{Graham} et~al.(2001){Graham}, {Erwin}, {Caon} \&
  {Trujillo}]{Graham2001}
{Graham} A.~W., {Erwin} P., {Caon} N., {Trujillo} I., 2001, \apjl, 563, L11

\bibitem[{G{\"u}ltekin} et~al.(2009){G{\"u}ltekin}, {Richstone}, {Gebhardt}
  et~al.]{Gultekin2009}
{G{\"u}ltekin} K., et~al., 2009, \apj, 698, 198

\bibitem[{Habouzit} et~al.(2017){Habouzit}, {Volonteri} \&
  {Dubois}]{Habouzit2017}
{Habouzit} M., {Volonteri} M., {Dubois} Y., 2017, \mnras, 468, 4, 3935

\bibitem[{Habouzit} et~al.(2016){Habouzit}, {Volonteri}, {Latif}, {Dubois} \&
  {Peirani}]{Habouzit2016}
{Habouzit} M., {Volonteri} M., {Latif} M., {Dubois} Y., {Peirani} S., 2016,
  \mnras, 463, 529

\bibitem[{Haehnelt} \& {Rees}(1993)]{HaehneltRees1993}
{Haehnelt} M.~G., {Rees} M.~J., 1993, \mnras, 263, 168

\bibitem[{H{\"a}ring} \& {Rix}(2004)]{HaringRix2004}
{H{\"a}ring} N., {Rix} H.-W., 2004, \apjl, 604, L89

\bibitem[{Heger} et~al.(2003){Heger}, {Fryer}, {Woosley}, {Langer} \&
  {Hartmann}]{Heger2003}
{Heger} A., {Fryer} C.~L., {Woosley} S.~E., {Langer} N., {Hartmann} D.~H.,
  2003, \apj, 591, 288

\bibitem[{Heger} \& {Woosley}(2010)]{HegerWoosley2010}
{Heger} A., {Woosley} S.~E., 2010, \apj, 724, 341

\bibitem[{Hinshaw} et~al.(2013){Hinshaw}, {Larson}, {Komatsu}
  et~al.]{Hinshaw2013}
{Hinshaw} G., et~al., 2013, \apjs, 208, 19

\bibitem[{Hirano} et~al.(2014){Hirano}, {Hosokawa}, {Yoshida}
  et~al.]{Hirano2014}
{Hirano} S., {Hosokawa} T., {Yoshida} N., {Umeda} H., {Omukai} K., {Chiaki} G.,
  {Yorke} H.~W., 2014, \apj, 781, 60

\bibitem[{Hirano} et~al.(2017){Hirano}, {Hosokawa}, {Yoshida} \&
  {Kuiper}]{Hirano2017}
{Hirano} S., {Hosokawa} T., {Yoshida} N., {Kuiper} R., 2017, Science, 357, 1375

\bibitem[{Hirschmann} et~al.(2014){Hirschmann}, {Dolag}, {Saro}, {Bachmann},
  {Borgani} \& {Burkert}]{Hirschmann2014}
{Hirschmann} M., {Dolag} K., {Saro} A., {Bachmann} L., {Borgani} S., {Burkert}
  A., 2014, \mnras, 442, 2304

\bibitem[{Hopkins} et~al.(2007){Hopkins}, {Richards} \&
  {Hernquist}]{Hopkins2007}
{Hopkins} P.~F., {Richards} G.~T., {Hernquist} L., 2007, \apj, 654, 731

\bibitem[{Jiang} et~al.(2009){Jiang}, {Fan}, {Bian} et~al.]{Jiang2009}
{Jiang} L., et~al., 2009, AJ, 138, 305

\bibitem[{Jiang} et~al.(2016){Jiang}, {McGreer}, {Fan} et~al.]{Jiang2016}
{Jiang} L., et~al., 2016, \apj, 833, 222

\bibitem[{Karakas}(2010)]{Karakas2010}
{Karakas} A.~I., 2010, \mnras, 403, 1413

\bibitem[{Karlsson} et~al.(2013){Karlsson}, {Bromm} \&
  {Bland-Hawthorn}]{Karlsson2013}
{Karlsson} T., {Bromm} V., {Bland-Hawthorn} J., 2013, Reviews of Modern
  Physics, 85, 809

\bibitem[{Katz} et~al.(2015){Katz}, {Sijacki} \& {Haehnelt}]{Katz2015}
{Katz} H., {Sijacki} D., {Haehnelt} M.~G., 2015, \mnras, 451, 2352

\bibitem[{Kelley} et~al.(2017){Kelley}, {Blecha}, {Hernquist}, {Sesana} \&
  {Taylor}]{Kelley2017}
{Kelley} L.~Z., {Blecha} L., {Hernquist} L., {Sesana} A., {Taylor} S.~R., 2017,
  \mnras, 471, 4508

\bibitem[{Klein} et~al.(2016){Klein}, {Barausse}, {Sesana} et~al.]{Klein2016}
{Klein} A., et~al., 2016, Phys. Rev. D, 93, 2, 024003

\bibitem[{Kormendy} \& {Ho}(2013)]{KormendyHo2013}
{Kormendy} J., {Ho} L.~C., 2013, ARA\&A, 51, 511

\bibitem[{Kormendy} \& {Richstone}(1995)]{KormendyRichstone1995}
{Kormendy} J., {Richstone} D., 1995, ARA\&A, 33, 581

\bibitem[{Latif} et~al.(2016){Latif}, {Omukai}, {Habouzit}, {Schleicher} \&
  {Volonteri}]{Latif2016}
{Latif} M.~A., {Omukai} K., {Habouzit} M., {Schleicher} D.~R.~G., {Volonteri}
  M., 2016, \apj, 823, 40

\bibitem[{Lodato} \& {Natarajan}(2006)]{LodatoNatarajan2006}
{Lodato} G., {Natarajan} P., 2006, \mnras, 371, 1813

\bibitem[{Loeb} \& {Rasio}(1994{\natexlab{a}})]{LoebRasio1994}
{Loeb} A., {Rasio} F.~A., 1994{\natexlab{a}}, \apj, 432, 52

\bibitem[{Loeb} \& {Rasio}(1994{\natexlab{b}})]{Loeb1994}
{Loeb} A., {Rasio} F.~A., 1994{\natexlab{b}}, \apj, 432, 52

\bibitem[{Madau} \& {Rees}(2001)]{MadauRees2001}
{Madau} P., {Rees} M.~J., 2001, \apjl, 551, L27

\bibitem[{Magorrian} et~al.(1998){Magorrian}, {Tremaine}, {Richstone}
  et~al.]{Magorrian1998}
{Magorrian} J., et~al., 1998, AJ, 115, 2285

\bibitem[{McAlpine} et~al.(2017){McAlpine}, {Bower}, {Harrison}
  et~al.]{McAlpine2017}
{McAlpine} S., {Bower} R.~G., {Harrison} C.~M., {Crain} R.~A., {Schaller} M.,
  {Schaye} J., {Theuns} T., 2017, \mnras, 468, 3395

\bibitem[{McConnell} \& {Ma}(2013)]{McConnellMa2013}
{McConnell} N.~J., {Ma} C.-P., 2013, \apj, 764, 184

\bibitem[{Merloni} et~al.(2012){Merloni}, {Predehl}, {Becker}
  et~al.]{eROSITA2012}
{Merloni} A., et~al., 2012,  arXiv:1209.3114

\bibitem[{Middleton} et~al.(2016){Middleton}, {Del Pozzo}, {Farr}, {Sesana} \&
  {Vecchio}]{Middleton2016}
{Middleton} H., {Del Pozzo} W., {Farr} W.~M., {Sesana} A., {Vecchio} A., 2016,
  \mnras, 455, L72

\bibitem[{Mortlock} et~al.(2011){Mortlock}, {Warren}, {Venemans}
  et~al.]{Mortlock2011}
{Mortlock} D.~J., et~al., 2011, Nature, 474, 616

\bibitem[{Nandra} et~al.(2013){Nandra}, {Barret}, {Barcons} et~al.]{ATHENA2013}
{Nandra} K., et~al., 2013, arXiv: 1306.2307

\bibitem[{Natarajan} et~al.(2017){Natarajan}, {Pacucci}, {Ferrara}
  et~al.]{Natarajan2017}
{Natarajan} P., {Pacucci} F., {Ferrara} A., {Agarwal} B., {Ricarte} A.,
  {Zackrisson} E., {Cappelluti} N., 2017, \apj, 838, 2, 117

\bibitem[{Natarajan} \& {Volonteri}(2012)]{NatarajanVolonteri2012}
{Natarajan} P., {Volonteri} M., 2012, \mnras, 422, 2051

\bibitem[{Nelson} et~al.(2015){Nelson}, {Pillepich}, {Genel}
  et~al.]{Nelson2015}
{Nelson} D., et~al., 2015, Astronomy and Computing, 13, 12

\bibitem[{Omukai} et~al.(2008){Omukai}, {Schneider} \& {Haiman}]{Omukai2008}
{Omukai} K., {Schneider} R., {Haiman} Z., 2008, \apj, 686, 801

\bibitem[{Pacucci} et~al.(2015){Pacucci}, {Ferrara}, {Volonteri} \&
  {Dubus}]{Pacucci2015}
{Pacucci} F., {Ferrara} A., {Volonteri} M., {Dubus} G., 2015, \mnras, 454, 4,
  3771

\bibitem[{Plowman} et~al.(2010){Plowman}, {Jacobs}, {Hellings}, {Larson} \&
  {Tsuruta}]{Plowman2010}
{Plowman} J.~E., {Jacobs} D.~C., {Hellings} R.~W., {Larson} S.~L., {Tsuruta}
  S., 2010, \mnras, 401, 4, 2706

\bibitem[{Portegies Zwart} \& {McMillan}(2002)]{PortegiesZwart2002}
{Portegies Zwart} S.~F., {McMillan} S.~L.~W., 2002, \apj, 576, 899

\bibitem[{Portinari} et~al.(1998){Portinari}, {Chiosi} \&
  {Bressan}]{Portinari1998}
{Portinari} L., {Chiosi} C., {Bressan} A., 1998, A\&A, 334, 505

\bibitem[{Reardon} et~al.(2016){Reardon}, {Hobbs}, {Coles} et~al.]{Reardon2016}
{Reardon} D.~J., et~al., 2016, \mnras, 455, 2, 1751

\bibitem[{Rees}(1984)]{Rees1984}
{Rees} M.~J., 1984, ARA\&A, 22, 471

\bibitem[{Regan} \& {Haehnelt}(2009)]{ReganHaehnelt2009}
{Regan} J.~A., {Haehnelt} M.~G., 2009, \mnras, 396, 343

\bibitem[{Regan} et~al.(2017){Regan}, {Visbal}, {Wise}, {Haiman}, {Johansson}
  \& {Bryan}]{Regan2017}
{Regan} J.~A., {Visbal} E., {Wise} J.~H., {Haiman} Z., {Johansson} P.~H.,
  {Bryan} G.~L., 2017, Nature Astronomy, 1, 0075

\bibitem[{Ricarte} \& {Natarajan}(2018)]{Ricarte2018}
{Ricarte} A., {Natarajan} P., 2018, \mnras, 481, 3278

\bibitem[{Rodriguez-Gomez} et~al.(2015){Rodriguez-Gomez}, {Genel},
  {Vogelsberger} et~al.]{Rodriguez-Gomez2015}
{Rodriguez-Gomez} V., et~al., 2015, \mnras, 449, 49

\bibitem[{Salcido} et~al.(2016){Salcido}, {Bower}, {Theuns}
  et~al.]{Salcido2016}
{Salcido} J., {Bower} R.~G., {Theuns} T., {McAlpine} S., {Schaller} M., {Crain}
  R.~A., {Schaye} J., {Regan} J., 2016, \mnras, 463, 1, 870

\bibitem[{Schauer} et~al.(2017){Schauer}, {Regan}, {Glover} \&
  {Klessen}]{Schauer2017}
{Schauer} A.~T.~P., {Regan} J., {Glover} S.~C.~O., {Klessen} R.~S., 2017,
  \mnras, 471, 4878

\bibitem[{Schaye} et~al.(2015){Schaye}, {Crain}, {Bower} et~al.]{Schaye2015}
{Schaye} J., et~al., 2015, \mnras, 446, 521

\bibitem[{Sesana} et~al.(2011){Sesana}, {Gair}, {Berti} \&
  {Volonteri}]{Sesana2011}
{Sesana} A., {Gair} J., {Berti} E., {Volonteri} M., 2011, Phys. Rev. D, 83, 4,
  044036

\bibitem[{Sesana} et~al.(2004){Sesana}, {Haardt}, {Madau} \&
  {Volonteri}]{Sesana2004}
{Sesana} A., {Haardt} F., {Madau} P., {Volonteri} M., 2004, \apj, 611, 2, 623

\bibitem[{Sesana} et~al.(2007){Sesana}, {Volonteri} \& {Haardt}]{Sesana2007}
{Sesana} A., {Volonteri} M., {Haardt} F., 2007, \mnras, 377, 4, 1711

\bibitem[{Sijacki} et~al.(2007){Sijacki}, {Springel}, {di Matteo} \&
  {Hernquist}]{Sijacki2007}
{Sijacki} D., {Springel} V., {di Matteo} T., {Hernquist} L., 2007, \mnras, 380,
  877

\bibitem[{Sijacki} et~al.(2009){Sijacki}, {Springel} \&
  {Haehnelt}]{Sijacki2009}
{Sijacki} D., {Springel} V., {Haehnelt} M.~G., 2009, \mnras, 400, 100

\bibitem[{Sijacki} et~al.(2015){Sijacki}, {Vogelsberger}, {Genel}
  et~al.]{Sijacki2015}
{Sijacki} D., {Vogelsberger} M., {Genel} S., {Springel} V., {Torrey} P.,
  {Snyder} G.~F., {Nelson} D., {Hernquist} L., 2015, \mnras, 452, 575

\bibitem[{Springel}(2010)]{Springel2010}
{Springel} V., 2010, \mnras, 401, 791

\bibitem[{Springel} et~al.(2005){Springel}, {White}, {Jenkins}
  et~al.]{Springel2005}
{Springel} V., et~al., 2005, Nature, 435, 629

\bibitem[{Springel} et~al.(2001){Springel}, {White}, {Tormen} \&
  {Kauffmann}]{Springel2001}
{Springel} V., {White} S.~D.~M., {Tormen} G., {Kauffmann} G., 2001, \mnras,
  328, 726

\bibitem[{Tanaka} \& {Haiman}(2009)]{TanakaHaiman2009}
{Tanaka} T., {Haiman} Z., 2009, \apj, 696, 2, 1798

\bibitem[{Thielemann} et~al.(2003){Thielemann}, {Argast}, {Brachwitz}
  et~al.]{Thielemann2003}
{Thielemann} F.~K., et~al., 2003, in { From Twilight to Highlight: The Physics
  of Supernovae\/}, edited by W.~{Hillebrandt}, B.~{Leibundgut},  331

\bibitem[{Tremaine} et~al.(2002){Tremaine}, {Gebhardt}, {Bender}
  et~al.]{Tremaine2002}
{Tremaine} S., et~al., 2002, \apj, 574, 740

\bibitem[{Valiante} et~al.(2018){Valiante}, {Schneider}, {Zappacosta},
  {Graziani}, {Pezzulli} \& {Volonteri}]{Valiante2018}
{Valiante} R., {Schneider} R., {Zappacosta} L., {Graziani} L., {Pezzulli} E.,
  {Volonteri} M., 2018, \mnras, 476, 1, 407

\bibitem[{Verbiest} et~al.(2016){Verbiest}, {Lentati}, {Hobbs}
  et~al.]{Verbiest2016}
{Verbiest} J.~P.~W., et~al., 2016, \mnras, 458, 1267

\bibitem[{Vogelsberger} et~al.(2013){Vogelsberger}, {Genel}, {Sijacki},
  {Torrey}, {Springel} \& {Hernquist}]{Vogelsberger2013}
{Vogelsberger} M., {Genel} S., {Sijacki} D., {Torrey} P., {Springel} V.,
  {Hernquist} L., 2013, \mnras, 436, 3031

\bibitem[{Vogelsberger} et~al.(2014{\natexlab{a}}){Vogelsberger}, {Genel},
  {Springel} et~al.]{Vogelsberger2014a}
{Vogelsberger} M., et~al., 2014{\natexlab{a}}, \mnras, 444, 1518

\bibitem[{Vogelsberger} et~al.(2014{\natexlab{b}}){Vogelsberger}, {Genel},
  {Springel} et~al.]{Vogelsberger2014b}
{Vogelsberger} M., et~al., 2014{\natexlab{b}}, \nat, 509, 177

\bibitem[{Volonteri} et~al.(2003){Volonteri}, {Haardt} \&
  {Madau}]{Volonteri2003}
{Volonteri} M., {Haardt} F., {Madau} P., 2003, \apj, 582, 559

\bibitem[{Volonteri} et~al.(2008){Volonteri}, {Lodato} \&
  {Natarajan}]{VolonteriLodatoNatarajan2008}
{Volonteri} M., {Lodato} G., {Natarajan} P., 2008, \mnras, 383, 1079

\bibitem[{Wang} et~al.(2019){Wang}, {Taylor}, {Federrath} \&
  {Kobayashi}]{Wang2019}
{Wang} E.~X., {Taylor} P., {Federrath} C., {Kobayashi} C., 2019, \mnras, 483,
  4, 4640

\bibitem[{Weinberger} et~al.(2018){Weinberger}, {Springel}, {Pakmor}
  et~al.]{Weinberger2018}
{Weinberger} R., et~al., 2018, \mnras, 479, 4056

\bibitem[{Whalen} \& {Fryer}(2012)]{WhalenFryer2012}
{Whalen} D.~J., {Fryer} C.~L., 2012, \apj, 756, L19

\bibitem[{Wise} et~al.(2019){Wise}, {Regan}, {O'Shea}, {Norman}, {Downes} \&
  {Xu}]{Wise2019}
{Wise} J.~H., {Regan} J.~A., {O'Shea} B.~W., {Norman} M.~L., {Downes} T.~P.,
  {Xu} H., 2019, \nat, 566, 7742, 85

\bibitem[{Woods} et~al.(2019){Woods}, {Agarwal}, {Bromm} et~al.]{Woods2019}
{Woods} T.~E., et~al., 2019, PASA, 36, 27

\bibitem[{Woosley} \& {Weaver}(1995)]{WoosleyWeaver1995}
{Woosley} S.~E., {Weaver} T.~A., 1995, The Astrophysical Journal Supplement
  Series, 101, 181

\end{thebibliography}

\end{document}